\documentclass[11pt]{article}
\usepackage{natbib}
\usepackage{diagbox}
\usepackage{amsmath,bm}
\usepackage{float}
\usepackage{xcolor}
\usepackage{subcaption}
\usepackage{soul}
\usepackage{authblk}
\usepackage[symbol]{footmisc}

\usepackage{amsmath,amsfonts,stmaryrd,amssymb} 

\usepackage{enumerate} 

\usepackage[ruled]{algorithm2e} 

\usepackage[framemethod=tikz]{mdframed} 

\usepackage{listings} 
\usepackage{geometry} 

\usepackage[utf8]{inputenc} 
\usepackage[T1]{fontenc} 
\usepackage{times}
\usepackage{cite}

\mdfdefinestyle{question}{
	innertopmargin=1.2\baselineskip,
	innerbottommargin=0.8\baselineskip,
	roundcorner=5pt,
	nobreak,
	singleextra={%
		\draw(P-|O)node[xshift=1em,anchor=west,fill=white,draw,rounded corners=5pt]{%
		Question \theQuestion\questionTitle};
	},
}

\newcounter{Question} 

\mdfdefinestyle{warning}{
	topline=false, bottomline=false,
	leftline=false, rightline=false,
	nobreak,
	singleextra={%
		\draw(P-|O)++(-0.5em,0)node(tmp1){};
		\draw(P-|O)++(0.5em,0)node(tmp2){};
		\fill[black,rotate around={45:(P-|O)}](tmp1)rectangle(tmp2);
		\node at(P-|O){\color{white}\scriptsize\bf !};
		\draw[very thick](P-|O)++(0,-1em)--(O);
	}
}

\title{Sensitivity analysis for nonignorable missing values in blended analysis framework: a study on the effect of bariatric surgery via electronic health records}
\date{}

\author[1]{Jungwun Lee}
\author[2]{Sebastien Haneuse}
\author[3]{Rajarshi Mukherjee}
\author[4, 5]{Tanayott Thaweethai\footnote{Corresponding author.}}

\affil[1]{Department of Biostatistics, Boston University School of Public Health, Boston, MA, USA}
\affil[2]{Department of Biostatistics, Harvard T.H. Chan School of Public Health, Boston, MA, USA}
\affil[3]{Department of Biostatistics, Harvard T.H. Chan School of Public Health, Boston, MA, USA}
\affil[4]{Massachusetts General Hospital, Boston, MA, USA}
\affil[5]{Department of Medicine, Harvard Medical School, Boston, MA, USA}

\begin{document}
\maketitle 

\begin{abstract}
This paper establishes a series of sensitivity analyses to investigate the impact of missing values in the electronic health records (EHR) that are possibly missing not at random (MNAR). EHRs have gained tremendous interest due to their cost-effectiveness, but their employment for research involves numerous challenges, such as selection bias due to missing data. The blended analysis has been suggested to overcome such challenges, which decomposes the data provenance into a sequence of sub-mechanisms and uses a combination of inverse-probability weighting (IPW) and multiple imputation (MI) under missing at random assumption (MAR). In this paper, we expand the blended analysis under the MNAR assumption and present a sensitivity analysis framework to investigate the effect of MNAR missing values on the analysis results. We illustrate the performance of my proposed framework via numerical studies and conclude with strategies for interpreting the results of sensitivity analyses. In addition, we present an application of our framework to the DURABLE data set, an EHR from a study examining long-term outcomes of patients who underwent bariatric surgery.

\end{abstract}

\section{Introduction}\label{intro}

\subsection{Electronic health records and their challenges}\label{EHR}

Electronic health records (EHR) consist of extensive patient-level information that is collected via encounters individuals have with healthcare systems, and include data regarding medical history,  hospital/emergency room visits, and clinical measurements. Such information is recorded primarily for clinical management of patients and/or billing purposes. Because EHR are naturally collected during all healthcare interactions, they are promising cost-effective resources for observational studies in public health and medical research as an alternative to conventional randomized clinical trials, which may be expensive, time-consuming, or ethically unjustified \citep{haneuse2016general}. 

Unfortunately, analyses using EHR are subject to challenges that may threaten the validity of conventional statistical analyses. As all EHR-based studies are inherently observational, patient-level treatments are not randomly assigned; thus, confounding bias can be a potential problem. In addition, not all clinical information relevant to a research question is routinely or consistently collected. For example, some patients may have missing baseline treatment, missing clinical information, missing laboratory measurements during follow-up, or have been disenrolled from the health plan before the end of the planned follow-up. Although not disenrolled, some patients may not have measurement histories in a certain period because they did not visit the healthcare system during that time. If such inconsistencies occur in variables of interest, records with missing values cannot be included in the analysis if the analysis usually requires complete records in all related variables.

Such missing values may cause serious biases in standard statistical analysis if inappropriately treated. For example, the complete-case analysis (CCA), employs subjects with complete data for the variables in an analysis model only and excludes subjects with incomplete values. However, it is based on a strong assumption in that subjects with complete records are systemically homogeneous with those who are excluded, and thus, they properly represent a target population. Such an assumption can be violated if the distribution of a variable in the analysis model among complete cases is systemically different from that of excluded ones. In addition, in the EHR context, a subject being a complete case largely hinges on an analysis model of interest; some patients without heart rate measurement may still be considered complete if an analysis model focuses on associations between weight loss and the bariatric surgical type and does not require heart rate. Such a unique nature of EHR yields the occurrence of incomplete record more non-accidental, making the CCA even less plausible.

Numerous alternatives have been proposed, including inverse probability weighting (IPW)\citep{Seaman2012} or multiple imputation (MI)\citep{rubin2018multiple} techniques. IPW employs subjects with complete subjects only while removing bias by weighting subjects with the inverse of their probability of being a complete case. MI is a simulation-based procedure which replaces each missing value with a set of plausible values, creating multiple imputed data sets. Analytic results are obtained by combining the results of standard complete-data analyses across the multiple imputed data sets. However, such conventional approaches hinges on an oversimplified assumption in that a single missing data mechanism can explain the missingness on a study sample, and thus may be insufficient to account for complex interplays between patients and decision makings that lead to a realization of a record \citep{haneuse2016general}. 

The blended analysis framework was proposed to reflect the unique data-generating process of EHR on the analysis model and provide broader flexibilities in addressing missing value problems in EHR-based studies. \citep{haneuse2016general} defined a term ``modularization", which decomposes the data-generating mechanism of EHR into a sequence of binary random variables, where each random variable denotes ``sub-mechanism" that describes an event that lead to a variable being ``observed", or a decision making that may affect the following sub-mechanisms \citep{haneuse2016general, thaweethai2020statistical}. The key idea is to decompose the data provenance into a sequence of multiple sub-mechanisms, where each sub-mechanism is a binary indicator that describes whether a specific decision or measurement of required data elements occurs. Such decomposition of the data provenance is referred to as ``modularization" \citep{haneuse2016general, peskoe2021adjusting, thaweethai2020statistical}. For a given data set and an analysis model, a modularization yields a sequence of sub-mechanisms, and each sub-mechanism is modeled either IPW or MI, based on the analyst's decision. After all sub-mechanisms are addressed, each imputed data set is weighted and analyzed, and the combination of results from multiple imputed data sets become the final analysis result \citep{thaweethai2020statistical}. 

\subsection{Sensitivity analysis on missing values}\label{SA}

\citet{little2019statistical} classified the missing data mechanism into three categories: missing completely at random (MCAR), missing at random (MAR), and missing not at random (MNAR). The missing data mechanism is MCAR if unobserved values on a variable are unrelated to the other observed variables and unobserved values themselves. Under MCAR, the sub-sample with complete data are statistically representative of the population from which they are derived, which is usually unrealistic in most settings. Under MAR, the missing data mechanism depends only upon the observed study variables. Under MAR, the probability of missingness does not depend on the missing data conditional on the observed variables. The missing data mechanism is MNAR if the missing data mechanism depends on the unobserved data, even after conditioning on the observed data. 

This paper is motivated by MNAR missing values in EHR. In EHR-based studies, the MAR assumption may not hold because missing values in EHR often appear in the complex interplays between patients, healthcare providers, and intractable factors. For example, some patients' weight measurements may no longer be available if they are satisfied with their weight loss after the bariatric surgery and thus disenroll from the healthcare system. On the contrary, some patients' weight measurements are not available because their condition worsens or they experience some side effects that need alternative treatment, and thus, they miss their healthcare visits. Since such MNAR missing values may yield biased and inefficient estimations if a standard MAR-based method is used, it is essential to investigate the validity of MAR assumption on the missing data mechanism. Unfortunately, the MAR assumption cannot be verified from the observed data, and an analyst can never rule out the possibility of MNAR \citep{schafer1997analysis}. Although including many variables in the imputation or selection model may decrease the possibility of MNAR \citep{schafer1997analysis}, ruling out the possibility of MNAR missing values is likely incorrect. Consequently, sensitivity analyses exploring the implications of departures from the MAR assumption are important to assess the robustness of the analysis results \citep{national2010panel,little2012prevention}.

Sensitivity analysis can be performed in various ways depending on how the external assumptions of an analyst are reflected in the analysis model. A basic strategy is the $\delta$-adjustment procedure, which shifts the distribution of imputed values or inverse-probability weights by a sensitivity parameter $\delta$ \citep{rubin1977formalizing}, where the sensitivity parameter reflects the analyst-oriented assumptions on distributions of unobserved data. Such a framework can be further extended to pattern-mixture models \citep{little1993pattern, ratitch2013missing} and tipping point analysis \citep{yan2009missing}. When adopted in the MI framework, the prespecified sensitivity parameter directly affects the imputed values for the missing variable via average imputed values \citep{mehrotra2017missing, leurent2018sensitivity}, correlation between observed and imputed data \citep{hsu2020multiple}, or weights of MAR-based estimates \citep{carpenter2007sensitivity}. In IPW, the inverse probabilities are calculated based on observed data and sensitivity parameter(s) so that the weights applied to the complete records become a function of both observed and prespecified sensitivity parameters. Both MI and IPW with $\delta$-adjustment offer a transparent and flexible approach to assess sensitivity to departures from MAR assumption. Recent examples can be found in \citep{carpenter2023multiple}. In the Bayesian framework, sensitivity parameters can be introduced as hyperparameters of prior distributions on parameters rather than as adjustments to the likelihood function. In this case, the prior distributions on population parameters reflect external assumptions on the missing data mechanism. See \citet{kaciroti2021bayesian} and its references for recent work.

This paper proposes a novel sensitivity analysis framework tailored for standard blended analysis that hinges on MAR assumptions. Specifically, we employ the $\delta$-adjustment strategy by modifying the selection models for IPW and the imputation models for MI as functions of both observed data and the missingness indicators combined with prespecified sensitivity parameters. Apart from traditional statistical models based on a single mechanism, our proposed method uses the $\delta$-adjustment method to modify sub-mechanisms defined via a blended analysis framework. Such modifications are applied to sub-mechanisms that are likely to be MNAR, and the sensitivity analysis is performed by varying the sensitivity parameter values and investigating the magnitude of the fluctuation of results by sensitivity parameters. Our proposed sensitivity analysis framework introduces sensitivity parameters in sub-mechanism levels and provides novel insights on investigating the robustness of analysis results to the departures of a missing data mechanism from MAR. Such an extension from a single mechanism to multiple mechanisms aligns with the development of the blended analysis from the traditional single mechanism-based method, which is based on the philosophy that a single missing data mechanism is insufficient to explain EHR's complex data provenance.

The rest of the paper is as follows. Section \ref{BAreview} reviews the modularization and blended analysis framework \citep{haneuse2016general, thaweethai2020statistical}. Section \ref{SAframe} modifies the blended analysis to deal with incomplete data that are possibly MNAR, and introduces how the sensitivity analysis can be performed under blended analysis. Section \ref{Numerical_Studies} consists of the numerical studies that illustrate the performance of our proposed method. Section \ref{application} illustrates how we applied this method to a real dataset. We conclude with a discussion in Section \ref{Discussions}.

\section{Review of the blended analysis framework}\label{BAreview}

\subsection{Motivating example: a study on the effect of surgical type on weight loss using DURABLE data}\label{motivation}

We will consider the Duration of Bariatric Long-term Effects (DURABLE) study as a motivating example . DURABLE is a large, multi-center EHR-based study investigating the long-term relationship between bariatric surgery and body weight, hypertension, and kidney disease. The study includes a dataset derived from EHR, which includes all $45,955$ individuals who underwent bariatric surgeries between 2007 and 2015 at one of three Kaiser Permanente health systems: Kaiser Permanente Washington (WA), Northern California (NC), and Southern California (SC). In addition, the data set includes $1,635,897$ enrolled participants who were eligible for bariatric surgery at least once between 2007 and 2015, having a BMI of at least 35 (kg/$m^2$). 

Two research questions of interest are (i) whether the different bariatric surgical types yield different changes in weight loss, conditioning on other demographic and clinical factors, and (ii) whether differences in weight loss between surgical types differ for individuals with and without CKD before surgery. The outcome of interest is the percent total BMI change defined as $PC_5 = 100(BMI_5 - BMI_0)/BMI_0$. Here, $BMI_0$ (BMI at baseline) is defined as the most recent BMI measurement among measurements before 30 days, including the surgery date. Any multiple BMI measures are averaged, and no BMI measurements within the time window are defined as missing in $BMI_0$. Similarly, $BMI_5$ (the follow-up BMI) is defined as the average BMI measures between 4.5 and 5.5 years after the surgery date.

To answer the two research questions above, one may fit an ordinary regression model using $PC_5$ as the outcome variable, and using CKD status at baseline ($CKD_0$), surgical type, and other patient-level variables as predictors. Ideally, for a patient to be included in such an analysis, they must have complete baseline and follow-up BMI information and other complete baseline measurements and demographic factors. Consequently, the analysis model can be written as follows:

\begin{align}\label{Analysismodel}
PC_5 = \beta_0 + \beta_{1}RYGB + \beta_{2}CKD_{0} + \beta_{3}RYGB\times CKD_{0} + CCS_0\boldsymbol\beta_{CCS_0} + {\bf X}_c\boldsymbol\beta_c + \epsilon,\quad\epsilon\sim (0,\sigma^2),
\end{align}
where $RYGB$ is a binary indicator of surgical type (1 if RYGB and 0 otherwise), and ${\bf X}_c$ denotes a vector of complete demographic factors. Hereafter, we denote the regression coefficients of surgical type, CKD indicator, and their interaction terms as $\beta_{RYGB}$, $\beta_{CKD_{0}}$, $\beta_{RYGB\times CKD_{0}}$, respectively.

\subsection{Modularization and blended analysis}\label{modularization}

One unique characteristic of EHR is the data-generating process, which is the byproduct of patient and provider behaviors, the clinical care the patient receives, and hospital billing practices. This brings up the importance of understanding its provenance, which can be understood by answering questions such as (1) what data (either variable or subject) should be collected in the ideal situation? and (2) why are some data observed while other data are not? \citep{haneuse2016general}. The first question is specified by the analyst in relation to the study they are attempting to conduct, whereas the second question requires an understanding of the data provenance: a series of events or decisions that lead to observing the "complete data" that the analyst requires for their analysis. The blended analysis framework has been designed to provide an integrated solution for the two questions above \citep{haneuse2016general, thaweethai2020statistical}. The key idea is to decompose the data provenance into a sequence of multiple sub-mechanisms, where each sub-mechanism is a binary indicator that describes whether a specific decision or measurement of required data elements occurs. Such decomposition of the data provenance is referred to as ``modularization" \citep{haneuse2016general, peskoe2021adjusting, thaweethai2020statistical}. 

\begin{figure}
\centering
\includegraphics[height = 8cm]{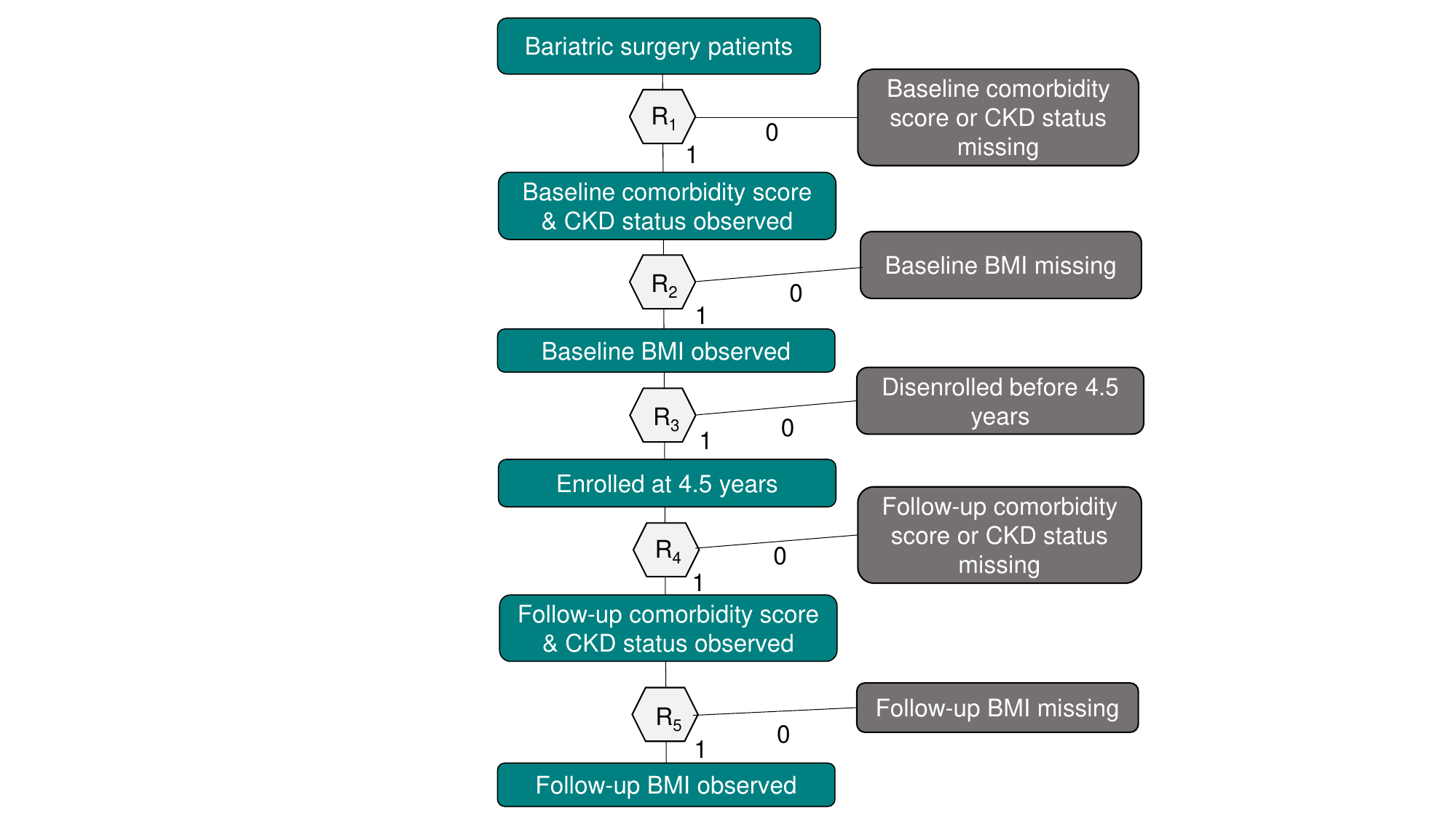}
\caption{A modularization which consists of five sub-mechanisms incorporating follow-up comoribity and CKD status in the 4th sub-mechanism in addition to the other variables needed to fit the analysis model.}\label{modularization_durable}
\end{figure}

Among the variables in the analysis model in Eq. (\ref{Analysismodel}), BMI measurements, CKD status, and the baseline CCS are subject to missing values. In addition, disenrollment from the healthcare system before their 5-year follow-up weight is measured can lead to missing data in the outcome as well. We address missing data problems in these variables by modularizing the study sample with five sub-mechanisms $[R_1,\ldots, R_5]$. Hereafter, we denote the $k$th sub-mechanism as $R_k$, which is a binary indicator with $R_{ik} = 1$ if the subject $i$ has a positive status related to the sub-mechanism $R_k$ and $R_{ik} = 0$ otherwise. Figure \ref{modularization_durable} illustrates how five sub-mechanisms explain the study data provenance. Specifically, $R_{i1} = 1$ if $i$th patient's baseline CKD and CCS values are observed and $0$ otherwise, and $R_{i2} = 1$ if $i$th patient's baseline BMI is observed and $0$ otherwise. $R_{i3}$ is a sub-mechanism for the enrollment status of $i$th patient in the healthcare systems at follow-up, and $R_{i4}$ indicates whether the $i$th patient's follow-up CKD status and CCS are both observed or not. Lastly, $R_{i5} = 1$ if $i$th patient's follow-up BMI is observed and $0$ otherwise.
 
\citet{peskoe2021adjusting} proposed using a sequence of IPW to account for modularized data provenance. For example, in Figure \ref{modularization_durable}, selection probabilities $P(R_1 = 1)$, $P(R_2 = 1\mid R_1 = 1)$,$\cdots$,$P(R_5 = 1\mid R_4 = R_3 = R_2 = R_1 = 1)$ are modeled separately, then the analysis is conducted with subjects with complete data weighted by the inverse of the product of five estimated probabilities. Further, \citet{thaweethai2020statistical} proposed the blended analysis framework, which employs both IPW and MI and thus creates its unique advantages over conventional single-mechanism approaches. 

The blended analysis framework has several advantages to traditional single-mechanism based methods. First, it increases the transparency of the missing data mechanism and the model assumptions for IPW or MI by decomposing the complex missing data mechanism into a sequence of sub-mechanisms. Considering the complex nature of EHR, it is a more reasonable approach than employing a single model in either IPW or MI. Second, an analyst can consider different link functions when building models for each sub-mechanism, and the blended analysis adds additional flexibility by blending MI or IPW appropriately. Also, blended analysis is highly scalable because each sub-mechanism is a binary indicator; an analyst may use a complicated model for IPW or imputation that reflects the data provenance. The inclusion of time-dependent covariates in modeling each sub-mechanism is also straightforward \citep{haneuse2016general,thaweethai2020statistical}. 

\subsection{Implementation}\label{BA_Implementation}

Let $D\in \mathbb{R}^{N\times P}$ be a collection of $P$ random variables from $N$ observations, and $\theta$ denotes the model parameter of interest. Some elements of $D$ are missing for some subjects. The first step is modularizing the data provenance by defining $K$ sub-mechanisms $[R_1,\ldots, R_K]$. Among $K$ sub-mechanisms, we denote $\tilde{\mathcal{K}} \subseteq \{1,\ldots,K\}$ as the collection of sub-mechanisms related to observing one or more variables. For example, in Figure \ref{modularization_durable}, $\tilde{\mathcal{K}} = \{1, 2, 4, 5\}$ because all sub-mechanisms except for $R_3$ are related to observing the baseline and follow-up BMI, CCS, and CKD, while $R_{3}$ describes patient enrollment status in the healthcare system. Based on the sub-mechanisms, we decompose $D$ into $[D^o, D_1,\ldots,D_K]$, where $D^o$ denotes the data for completely observed variables and each $D_j$ denotes the data for incompletely observed variables that are related to $R_j$.

To implement the blended analysis, the analyst determines whether each sub-mechanism $R_k$ with $k\in \tilde{\mathcal{K}}$ will be addressed with IPW or MI. Note that a sub-mechanism with $k\notin \tilde{\mathcal{K}}$ can only be handled via IPW, as there is no missing data to impute.  See \citet{thaweethai2020statistical} for additional discussion on choosing IPW or MI for a sub-mechanism. Based on these decisions, we divide $[1,\ldots,K]$ into $[\mathcal{K}_{IPW}, \mathcal{K}_{MI}]$, where $\mathcal{K}_{IPW}$ and $\mathcal{K}_{MI}$ denotes indices for which IPW and MI are used, respectively. Lastly, each $D_k$ can be decomposed into $[D_k^o, D_k^m]$, where $D_k^o$ denotes completely observed elements and $D_k^m$ denotes missing values in $D_k$.

Based on the assignments $[\mathcal{K}_{IPW}, \mathcal{K}_{MI}]$, the blended analysis proceeds by performing either IPW and MI sequentially from $R_1$ to $R_K$. Specifically, if $k\in\mathcal{K}_{IPW}$, then the probabilities of $R_{k} = 1$ are estimated for all subjects using the observed data, and the subjects with $R_{k} = 0$ are excluded. If $k\in\mathcal{K}_{MI}$, an imputation model $h_k$ is estimated using subjects with $R_{k} = 1$, and $D_k^m$ is imputed once. In such a way, all $K$ sub-mechanisms can be addressed, and the entire process is repeated $M > 1$ times, providing $M$ imputed data sets with weights, where the weight is a product of weights from all sub-mechanisms that are addressed by the IPW. The analysis model is applied to $M$ datasets with weights, resulting in $M$ parameter estimates. The final result is obtained by combining $M$ results based on Rubin's rules \citep{rubin2018multiple}. 

The unbiased estimation of the blended analysis hinges on the correct specification of the selection/imputation models for all sub-mechanisms. Thus, all selection and imputation models should be carefully constructed based on the appropriate assumption on the conditional distribution of $D_k^m$ given observed data and also should reflect the analysis model. Such a condition opens a new challenge; if one or more sub-mechanisms are incorrectly modeled under MAR assumption while the true behavior is MNAR, the following blended analysis may yield biased estimates. Unfortunately, a MAR assumption on a sub-mechanism cannot be verified by observed data. Consequently, the sensitivity analysis that investigates the robustness of the analysis result toward the violation of the MAR assumption should be accompanied.

\section{The blended analysis under MNAR and sensitivity analysis}\label{SAframe}

In this section, we establish a protocol for sensitivity analyses under the blended analysis framework to assess how the analysis results is affected by potential violation of MAR assumptions on missing value mechanisms. The basic tool is the $\delta$-adjustment approach, where a prespecified sensitivity parameter $\delta$ reflects the magnitude of the MNAR missingness. In this paper, we propose implementing the sensitivity analysis at the sub-mechanism level using the $\delta$-adjustment method for a particular sub-mechanism, or sub-mechanisms, of interest.

We introduce a vector of sensitivity parameters $\boldsymbol\delta = [\delta_1,\ldots,\delta_K]$ where each $\delta_k$ corresponds to the sub-mechanism $R_k$, where each $\delta_k$ reflects t he association between the sub-mechanism $R_k$ and the unobserved data $D_K^{m}$ via the sensitivity function $\xi_k(R_k;\delta_k)$. Both the sensitivity functions and parameters must be prespecified by analyst. Next, for given $\boldsymbol\delta$, the blended analysis under MNAR can be outlined as follows:

\begin{enumerate}
\item Define $K$ sub-mechanisms $[R_1,\ldots,R_K]$ and assign $\{1,\ldots,K\}$ to $[\mathcal{K}_{IPW}, \mathcal{K}_{MI}]$. 
\item Specify ${\boldsymbol\delta} = [\delta_1,\ldots,\delta_k]$.
\item Address $R_1$ based on the assignment. Hereafter, $\hat{\psi}_{k,\delta_k}$ is the parameter estimates related to the selection model for $R_k$ subject to $\delta_k$. Also, ${\psi}_{k}$ is the parameter of the imputation model for $D_k^m$, missing values related to sub-mechanism $R_k$.
\begin{enumerate}
\item If $1\in \mathcal{K}_{IPW}$, estimate the selection probability and calculate $w_{i1}$ for all subjects with $R_{i1} = 1$ as follows.
\begin{align}\label{IPW_MNAR_1}
w_{i1,\delta} = P(R_{i1} = 1\mid D^o, D_1^o;\hat{\psi}_{1,\delta}^{(l)},\delta_1)^{-1}.
\end{align}
\item If $1\in \mathcal{K}_{MI}$, estimate an imputation model and draw $[D_{i1,\delta}^{m(1)},\ldots,D_{i1,\delta}^{m(M)}]$ for all subjects with $R_{1i} = 0$ as follows.
\begin{align}\label{MI_MNAR_1}
\psi_1^{*}&\overset{iid}{\sim} \pi_1(D_1^m\mid D^o,D_1^o,\ldots,D_K^o,R_{1} = 1),\nonumber\\
D_{i1}^{m(l)}&\overset{iid}{\sim} h_1(D_1^m\mid D^o,D_1^o,\ldots,D_K^o,R_{1};\psi_1^{*},\delta_1),~l=1,\ldots,M.
\end{align}
where $\pi_1(\psi_1\mid D^o,D_{1}^{(l)},\ldots,D_{k-1}^{(l)},D_{k}^o,\ldots,D_{K}^o,R_{1} = 1)$ is the posterior distribution of $\psi_k$ given observed data and $l$th imputed data set.
\end{enumerate}
\item Based on assignment $[\mathcal{K}_{IPW}, \mathcal{K}_{MI}]$, implement $R_2,\ldots,R_K$ sequentially.
\begin{enumerate}
\item If $k\in \mathcal{K}_{IPW}$, estimate selection probability and calculate $w_{ik}$ at given $\delta_k$ for all subjects with $\bar{R}_{ik} = 1$ as follows.
\begin{align}\label{IPW_MNAR_k}
w_{ik}^{(l)} = P(R_{ik} = 1\mid D^o,D_{1}^{(l)},\ldots,D_{k-1}^{(l)},{\bar R}_{k-1} = 1,D_k;\hat{\psi}_{k,\delta}^{(l)},\delta_k)^{-1},~l=1,\ldots,M.
\end{align}
\item If $k\in \mathcal{K}_{MI}$, estimate an imputation model and draw $[D_{ik,\delta}^{m(1)},\ldots,D_{ik,\delta}^{m(M)}]$ for all subjects with $R_{ik} = 0$ as follows.
\begin{align}\label{MI_MNAR_K}
\psi_{k}^{(l)*}\overset{iid}{\sim} \pi_k(\psi_{k}^{(l)}&\mid D^o,D_{1}^{(l)},\ldots,D_{k-1}^{(l)},D_{k}^o,\ldots,D_{K}^o, {\bar R}_{k-1} = 1, R_k = 1),~l=1,\ldots,M.\nonumber\\
D_{ik,\delta}^{m(l)}\overset{iid}{\sim} h_k(D_k^m &\mid D^o,D_{1}^{(l)},\ldots,D_{k-1}^{(l)},D_{k}^o,\ldots,D_{K}^o, {\bar R}_{k-1} = 1, R_k;\psi_{k}^{(l)*},\delta_k),~l=1,\ldots,M.
\end{align}
where $\pi_k(\psi_k^{(l)}\mid D^o,D_{1}^{(l)},\ldots,D_{k-1}^{(l)},D_{k}^o,\ldots,D_{K}^o,{\bar R}_{k-1} = 1)$ is a posterior distribution of $\psi_k$ given observed and $l$th imputed data.
\end{enumerate}
\item For $l=1,\ldots,M$, denote $D^{(l)} = \{D^{o},D_1^{(l)},\ldots,D_K^{(l)}\}$ as the $l$th data set and $w_{i,\delta}^{(l)} = \prod\limits_{k=1}^Kw_{ik,\delta}^{(l)}$ as the corresponding weight for $D^{(l)}$. Fit the analysis model on $D^{(l)}$ using subjects with $R_K = 1$ for all $k\in \mathcal{K}_{IPW}$ and weighting by $w_{i,\delta}^{(l)}$, and obtain parameter estimates ${\hat\theta}_{\boldsymbol\delta}^{(l)}$. The blended analysis estimate of $\theta$ is ${\hat\theta}_{\boldsymbol\delta} = \frac{1}{M}\sum\limits_{l=1}^M{\hat\theta}_{\boldsymbol\delta}^{(l)}$.
\end{enumerate}

Under MNAR, both selection and imputation models are functions of external information via prespecified sensitivity parameter $\delta_k$, along with observed data. As shown in Eq. (\ref{IPW_MNAR_1}) through (\ref{MI_MNAR_K}), each $\delta_k$ plays different roles in IPW and MI. In IPW, the association between $D_k$ and its sub-mechanism $R_k$ is described via $\xi_k(D_k;\delta_k)$ and embedded in the selection probability. In MI, the imputation model parameter estimates $\hat{\psi}_{k}^{(l)}$ does not depend on $\delta_k$. Instead, we construct an imputation model by estimating $\hat{\psi}_{k}^{(l)}$ using only observed data and shift the imputation distribution using sensitivity function $\xi_k(R_k;\delta_k)$, while the correct estimation of the selection model parameters under MNAR requires accounting for both observed data and sensitivity parameter. 

When performing sensitivity analyses, the analyst specifies a set of sensitivity parameter values $\boldsymbol{\delta}$, implement the blended analysis over different values of $\boldsymbol{\delta}$, and check the fluctuations of $\hat\theta_{\delta}$. For each sub-mechanism $R_k$, the analyst should specify a plausible range of $\delta_k$ that reflects the professional domain knowledge as well as the goal of sensitivity analysis, including $0$, a value that implies MAR assumption. The large fluctuation of ${\hat\theta}_{\boldsymbol{\delta}}$ over different sensitivity parameters $\delta_k$ implies that the analysis result is sensitive to the choice of $\delta_k$. In particular, if $\hat\theta_{\boldsymbol{\delta}}$ largely fluctuates when $\delta_k$ is close to $0$, this implies that the analysis result highly depends on the assumption that sub-mechanism $R_k$ is MAR. 

While exploring the values of $\hat\theta_{\boldsymbol{\delta}}$ over a single $\delta_k$ is straightforward and can be done using the $\delta_k$-$\hat\theta_{\boldsymbol{\delta}}$ plot, the investigation on $\hat\theta_{\delta}$ over multiple $\delta_k$ can be challenging. To overcome such difficulties, we start with conditional sensitivity analyses by varying a single $\delta_k$ and holding other sensitivity parameters to be $0$. If two or more sub-mechanisms present fluctuations, then we proceed to two-way sensitivity analyses by only considering sub-mechanisms with high fluctuations. We will revisit this topic with a real-data example in Sections \ref{Conditional_sensitivity_analysis} and \ref{Twoway_sensitivity_analysis}. In the rest of the section, we elaborate on how the selection and imputation models can be specified and discuss estimation strategies. 

\subsection{Sub-mechanisms using multiple imputation}\label{MI}

In this section, we discuss how the $\delta$-adjustment procedure is applied to implement multiple imputation under alternative MNAR assumptions. Suppose we assign $k\in \mathcal{K}_{MI}$. As introduced in Eq. (\ref{MI_MNAR_1}) and Eq. (\ref{MI_MNAR_K}), we sample a value from the imputation model $h_k$. Generating the $l$th imputed value $D_{ik}^{m(l)}$ involves sampling $\hat{\psi}_{k}^{(l)*}$ from its posterior distribution $\pi_k$ and sampling $D_{ik}^{m(l)}$ using observed data and a posterior sample $\hat{\psi}_{k}^{(l)*}$ as follows:

\begin{align}\label{MI_MNAR_K2}
\hat{\psi}_{k}^{(l)*}\sim \pi_k(\psi_{k}&\mid D^o,D_{1}^{(l)},\ldots,D_{k-1}^{(l)},D_{k}^o,\ldots,D_{K}^o, {\bar R}_{k-1} = 1), \nonumber\\
D_{ik,\delta}^{m(l)}\overset{iid}{\sim} h_k(D_k^m&\mid D^o,D_{1}^{(l)},\ldots,D_{k-1}^{(l)},D_{k}^o,\ldots,D_{K}^o, {\bar R}_{k-1} = 1, R_k;\hat{\psi}_{k}^{(l)*}, \delta_k).
\end{align}

Note that the posterior distribution $\pi_k$ is estimated based on observed data and does not depend on unobserved data $D_k^m$, $R_k$, and $\delta_k$. The contribution of $R_k$ and the sensitivity parameter $\delta_k$ on the imputed value $D_{ik,\delta_k}^{m(l)}$ appears as a form of sensitivity function $\xi_k(R_k;\delta_k)$. A sensitivity function should be defined so that $\xi_k(R_k;\delta_k)$ does not depend on $R_k$ at the specific $\delta_k$, which yields the MAR condition. When $k\in \mathcal{K}_{MI}$, we use $\xi_k(R_k;\delta_k) = \delta_k(1 - R_k)$ so that it does not depend on $R_k$ when $\delta_k = 0$ \citep{leacy2017analyses}. Under this condition, $\delta_k = 0$ implies that the imputations are implemented under MAR.

For example, consider a sub-mechanism $R_k$ with $k\in \mathcal{K}_{MI}$ and let $D_k$ be a continuous random variable. The imputed value $D_{ik,\delta}^{m(l)}$ for subject $i$ with $R_{ik} = 0$ at $l$th imputation can be generated as follows:

\begin{align}\label{example1}
[\psi_{k}^{(l)*},\sigma_{k,(l)}^{2*}] &\sim \pi_k(\psi_{k}\mid D^o,D_{1}^{(l)},\ldots,D_{k-1}^{(l)},D_{k}^o,\ldots,D_{K}^o, {\bar R}_{k-1} = 1), \nonumber\\
\mu_k^{(l)} &= g(D^o,D_{1}^{(l)},\ldots,D_{k-1}^{(l)},D_{k}^o,\ldots,D_{K}^o;\psi_{k}^{(l)*}) + \delta_k(1 - R_k), \nonumber\\
D_{ik,\delta}^{m(l)} &\overset{iid}{\sim} 
 N(\mu_k^{(l)}, \sigma_{k,(l)}^{2*}).
\end{align}

In this example, the posterior distribution of imputed values is a function of observed/imputed data and the sensitivity function $\xi_k(R_k;\delta_k) = \delta_k(1 - R_k)$, where $\delta_k$ represents the difference in the mean of $D_k$ for subjects with missing $D_k$ compared to the subjects with observed $D_k$. Intuitively, if one specifies $\delta_k > 0$, this reflects the assumption that the average missing values of $D_k$ are greater than the average observed values of $D_k$ by $\delta_k$ units. Note that $\delta_k = 0$ implies MAR assumption. If $D_k$ is a binary random variable, then the imputation model can be modified as follows:

\begin{align}\label{example2}
\psi_{k}^{(l)*} &\sim \pi_k(\psi_{k}\mid D^o,D_{1}^{(l)},\ldots,D_{k-1}^{(l)},D_{k}^o,\ldots,D_{K}^o, {\bar R}_{k-1} = 1), \nonumber\\
p_k^{(l)} &= expit(g(D^o,D_{1}^{(l)},\ldots,D_{k-1}^{(l)},D_{k}^o,\ldots,D_{K}^o;\psi_{k}^{(l)*}) + \delta_k(1 - R_k)), \nonumber\\
D_{ik,\delta}^{m(l)} &\sim Bin(1, p_k^{(l)}).
\end{align}

where $expit(x) = e^x/(1 + e^x)$. In Eq. (\ref{example2}), $\delta_k$ represents the difference in the log-odds of $D_k = 1$ for subjects with $R_{ik} = 0$ compared to subjects with $R_{ik} = 1$. Similar to Eq. (\ref{example1}), a positive value of $\delta_k$ implies that subjects with $R_{ik} = 0$ have higher odds of $D_k^m = 1$ compared to $D_k^m = 0$ than subjects with $R_{ik} = 1$. If $\delta_k = 0$, the imputation model only depends on observed/imputed data and thus implies the MAR assumption. 

\subsection{Sub-mechanisms using inverse-probability weighting}\label{IPW}

In this section, we describe how to estimate selection models as discussed in Eq. (\ref{IPW_MNAR_1}) and Eq. (\ref{IPW_MNAR_k}). Suppose we have $k\in \mathcal{K}_{IPW}$ and the previous $k - 1$ sub-mechanisms have been addressed via both MI and IPW. As an initial step, we specify the selection model as follows.

\begin{align}\label{IPW_MNAR3}
P(R_k = 1\mid D^o,D_{1}^{(l)},\ldots,D_{k-1}^{(l)},{\bar R}_{k-1} = 1) &= h_k(D^o,D_{1}^{(l)},\ldots,D_{k-1}^{(l)},{\bar R}_{k-1} = 1, D_k;\hat{\psi}_{k,\delta}^{(l)},\delta_k), \nonumber\\
logit(h_k(D^o,D_{1}^{(l)},\ldots,D_{k-1}^{(l)},{\bar R}_{k-1} = 1, D_k;\hat{\psi}_{k,\delta}^{(l)},\delta_k)) &= g(D^o,D_{1}^{(l)},\ldots,D_{k-1}^{(l)};\hat{\psi}_{k,\delta}^{(l)}) + \xi_k(D_k;\delta_k). 
\end{align}

In Eq. (\ref{IPW_MNAR3}), $logit(x) = log(x/(1 - x))$ denotes a logit link function, and $\psi_{k}(D_k;\delta_k)$ denotes the sensitivity function with prespecified $\delta_k$. Also, $g(D^o,D_{1}^{(l)},\ldots,D_{k-1}^{(l)};\psi_{k,\delta}^{(l)})$ is a link function that reflects the association between the sub-mechanism $R_k$ and observed/imputed data. The selection probability does not depend on $D_k$ if $\psi_{k}(D_k;\delta_k) = 0$ and the missing data mechanism on $R_k$ becomes MAR. However, when $\xi_k(D_k;\delta_k) \neq 0$, standard estimation cannot be applied due to missing values in $D_k$. Instead, we solve the estimating equation for $\xi_k(D_k;\delta_k)$ for a fixed $\delta_k$ as follows:

\begin{align}\label{IPW_MNAR4}
\sum\limits_{i=1}^{N}\frac{\partial h_k(D^o,D_{1}^{(l)},\ldots,D_{k-1}^{(l)}, D_k;\psi_{k,\delta}^{(l)}, \delta_k)}{\partial\psi_{k,\delta}^{(l)}}\bar{R}_{i(k - 1)}\left\{\frac{R_{ik}}{h_k(D^o,D_{1}^{(l)},\ldots,D_{k-1}^{(l)}, D_k;\psi_{k,\delta}^{(l)}, \delta_k)}-1\right\} = \boldsymbol{0} \in \mathbb{R}^{P}. 
\end{align}

In Eq. (\ref{IPW_MNAR4}), $P$ is the dimension of $\psi_{k,\delta}$, and thus Eq. (\ref{IPW_MNAR4}) is a system of $P$ equations. Solving Eq. (\ref{IPW_MNAR4}) does not involve missing values because $\bar{R}_{i(k - 1)} = 1$ implies missing values related to the previous sub-mechanisms are all addressed (weighted by IPW or imputed by MI), and also $R_{ik} = 0$ for all subjects with missing $D_k$ \citep{wen2018semi, su2022sensitivity}. Note that the parameter $\psi_{k,\delta}^{(l)}$ is determined by the prespecified $\delta_k$, because solving Eq. (\ref{IPW_MNAR4}) involves $\xi_k(D_k;\delta_k)$. 

Throughout this paper, we use $\xi_k(D_k;\delta_k) = \delta_kD_k$ for $k\in \mathcal{K}_{IPW}$ so that $\delta_k = 0$ implies MAR. Such $\xi_k(D_k;\delta_k)$ is also easy to understand; $\delta_k$ represents the increment of log-odds of $R_k = 1$ as $D_k$ increases by one unit. Consequently, a positive $\delta_k$ represents that subjects with small $D_k$ are more likely to have missing $D_k$, while subjects with larger $D_k$ are more likely to be observed, conditioning on other factors. For example, we can use logistic linear regression for Eq. (\ref{IPW_MNAR3}) so that $g(D^o,D_{1}^{(l)},\ldots,D_{k-1}^{(l)}, D_k;\psi_{k,\delta}^{(l)}) = \beta_{0,\delta}^{(l)} + \sum\limits_{j=1}^{k-1}\beta_{j,\delta}^{(l)}D_{j}^{(l)}$ and $\psi_{k,\delta}^{(l)} = [\beta_{0,\delta}^{(l)},\ldots,\beta_{k-1,\delta}^{(l)}]$. This provides the partial derivative $\frac{\partial h_k(D^o,D_{1}^{(l)},\ldots,D_{k-1}^{(l)}, D_k;\psi_{k,\delta}^{(l)}, \delta_k)}{\partial\psi_{k,\delta}^{(l)}} = [1,D_1^{(l)},\ldots,D_{k-1}^{(l)}]'$, and we can obtain $\hat{\psi}_{k,\delta}^{(l)}$ by solving $k$-dimensional estimating equation.

\subsection{Inference}\label{inference}

Rubin's combining rules have been widely used for standard errors of combined estimates from multiply imputed complete datasets under conventional MI \citep{rubin2018multiple}. \citet{Seaman2012} proposed using Rubin's rules for calculating standard errors of estimates when combining IPW with MI to address missing data problems. However, Rubin's rules may yield biased estimates if the imputation or selection models are misspecified \citep{Thaweethai2020Robust}. To avoid such a challenge, \citet{schomaker2018bootstrap} proposed using Bootstrap-MI, where MIs are applied to bootstrapped data sets. Specifically, we obtain $B$ bootstrapped data set generated from the raw (including missing) data. The blended analysis is applied for each bootstrapped dataset with a fixed vector of sensitivity parameters $\boldsymbol{\delta}$, and $B\times M$ estimates are produced if at least one sub-mechanism receives MI. If all sub-mechanisms receive IPW, then only $B$ estimates are obtained. Let $\hat\theta_{\delta}^{(m,b)}$ be the parameter estimates from $b$th bootstrapped and $m$th imputed data set. We combine $M$ estimates into a single point estimate as $\hat\theta_{\delta}^{(*,b)} =\frac{1}{M}\sum\limits_{m = 1}^M\hat\theta_{\delta}^{(m,b)}$ within each bootstrap sample and obtain $B$ estimates. The $100(1 - \alpha)\%$ CI for $\theta_{\delta}$ is obtained using $100(\alpha/2)\%$ and $100(1 -\alpha/2)\%$ percentiles of the ordered bootstrap estimates $\{\hat\theta_{\delta}^{(*,b)}; b=1\ldots,B\}$. We use $B = 300$ and $M = 10$ in both the simulation and the real-data studies.

\section{Numerical Studies}\label{Numerical_Studies}

In this section, we evaluate the performance of our proposed sensitivity analysis framework via simulation. This section has two goals: first, to illustrate that the blended analysis procedure provides valid estimation when the sensitivity function and parameters are correctly specified, and second, to examine the effect of misspecification. We consider several data-generating scenarios that mimic the motivating data example where data are MNAR. There are four variables: a baseline binary treatment $X$, time-dependent covariates $[Z_1, Z_2]$ where $Z_1$ is measured at baseline and $Z_2$ is measured at 8 months after baseline but before the measurement of outcome, and a continuous outcome variable $Y$ measured 1 year from baseline. Briefly, $Z_1$ and $Z_2$ are correlated, the value of $Z_2$ is affected by $X$, and the outcome $Y$ is affected by $X$ and $Z_2$. We are interested in fitting the following analysis model using ordinary linear regression:

\begin{align}\label{simul_analysismodel}
Y_i=\beta_0+\beta_1X_i+\beta_2Z_{i1}+\beta_3X_iZ_{i1}+\epsilon_i,\quad\epsilon_i\sim N(0,\sigma^2),\quad i=1,\ldots,1000.
\end{align}

\begin{figure}
\centering
\includegraphics[height = 8cm]{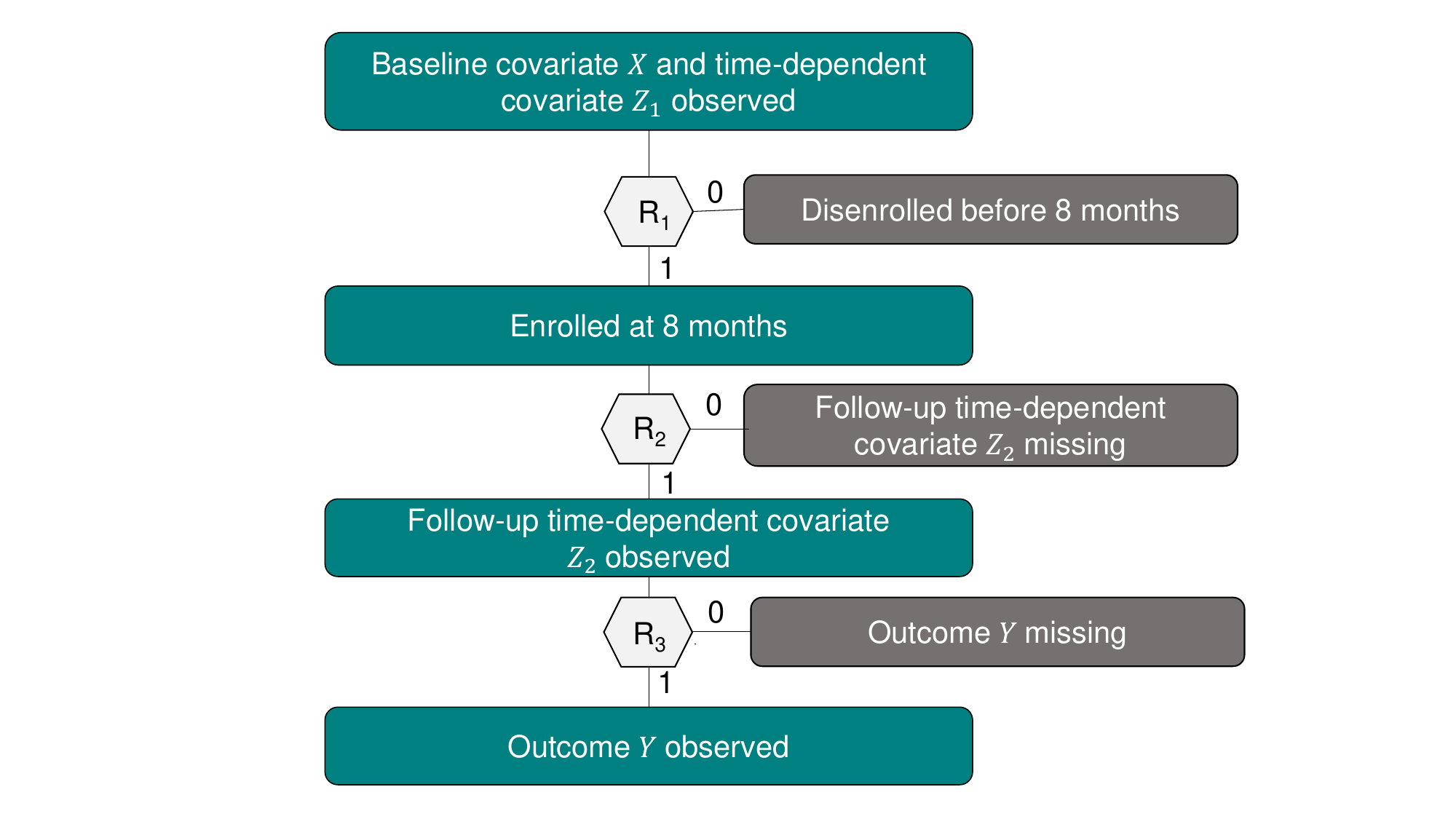}
\caption{A modularization of simulation data provenances with three sub-mechanisms. $R_1$, $R_2$, and $R_3$ represent individual sub-mechanisms on the path to observing $Z_2$ and $Y$.}\label{diag_simul}
\end{figure}

Baseline covariates $[X, Z_1]$ are fully observed, but $Z_2$ and $Y$ are subject to missingness due to two reasons: patients who disenroll from the study after 8 months of its initiation will have missing values in $Z_2$ and $Y$. In addition, patients still enrolled in the study may still have missing values in $Z_2$ and/or $Y$. We modularize the data provenance with three sub-mechanisms $[R_1, R_2, R_3]$. $R_{i1}$ represents whether a patient $i$ is enrolled in the study 8 months after initiation, $R_{i2}$ denotes whether $Z_2$ is measured, and $R_{i3}$ denotes whether $Y$ is measured (Figure \ref{diag_simul}).

Data are generated from the following distributions:
\begin{align}\label{simul_completedata}
Z_{i1} &\overset{iid}{\sim} Bin(1, expit(-0.5)), \nonumber\\
X_{i1} &\overset{iid}{\sim} Bin(1, expit(-0.5 - 0.25Z_{i1})), \nonumber\\
Z_{i2} &\overset{iid}{\sim} Bin(1, expit(-1.15 - 0.35X_{i1} + 0.6Z_{i1} + 0.4X_1Z_{i1})), \nonumber\\
Y_{i} &\overset{iid}{\sim} N(0.45 - 0.45X_i + 1.40Z_{i2} - 1.8X_iZ_{i2}, 0.3^2). 
\end{align}

The distribution of $Y$ is directly related to $[X, Z_2]$ only, and thus, its associations with $[X, Z_1, XZ_1]$ are implicit. Consequently, we empirically approximate the true values of analysis model parameters $\boldsymbol\beta$ in Eq. (\ref{simul_analysismodel}). The approximated analysis model parameters are $\boldsymbol\beta = [0.787, -0.859, 0.176, -0.253]$. In addition, we generate missingness on $[Z_2, Y]$ of each simulated data as 

\begin{align}\label{simul_missingdata}
\lambda(X, Z_1) &= exp(1.25 + 0.5X - 0.55Z_1 - 0.2XZ_1), \nonumber\\
T&\sim Weibull(\lambda(X, Z_1), 1), \nonumber\\
P(R_1 &= 1\mid X, Z_1) = P(T > 2/3\mid X, Z_1), \nonumber\\
P(R_2 &= 1\mid X, Z_1, Z_2) = expit(1.20 - 0.70X + 0.65Z_1 - 0.55XZ_1 + \delta_{2}Z_2), \nonumber\\
P(R_3 &= 1\mid X, Z_2, Y) = expit(0.45 + 0.35X - 0.70Z_2 - 0.65XZ_2 + \delta_{3}Y),
\end{align}
where $Weibull(\lambda(X, Z_1), 1)$ denotes the Weibull distribution with shape parameter $\lambda(X, Z_1)$ and the scale parameter 1, and $T$ denotes the disenrollment time. The hypothetical follow-up time is $8$ months, so patients with $T\leq 2/3$ are disenrolled before $Z_2$ or $Y$ are measured. In addition, the probabilities of $Z_2$ and $Y$ being unobserved are functions of $[X, Z_1, Z_2]$ and $[X, Z_2, Y]$, respectively. Note that the missing data mechanism is MAR if $\delta_2 = \delta_3 = 0$ and MNAR otherwise. When $\delta_2 = \delta_3 = 0$, distributions of three sub-mechanisms are $P(R_1 = 1)\approx 0.80$, $P(R_1 = R_2 = 1)\approx 0.57$, and $P(R_1 = R_2 = R_3 = 1)\approx 0.31$. Also, in this scenario, $\tilde{\mathcal{K}} = \{2, 3\}$ because $R_1$ is not related to observing a random variable. IPW (denoted as ``I") is the only possible method, while $[R_2, R_3]$ are open to both IPW or MI (denoted as ``M"). Consequently, there are four possible options for the blended analysis: ``III", ``IMI", ``IIM", and ``IMM", depending on how $[\mathcal{K}_{IPW}, \mathcal{K}_{MI}]$ is assigned. 

We consider two MNAR scenarios. First, we set $[\delta_2, \delta_3] = [0.5, 0]$ in Eq. (\ref{simul_missingdata}) so that $Z_2$ is MNAR (Section \ref{mnar_r22}). Second, we set $[\delta_2, \delta_3] = [0, 0.5]$ so that $Y$ is MNAR (Section \ref{mnar_r3}). For each simulation, we simulate $1,000$ data sets of sample size $1,000$. For each simulated dataset, we repeat steps $2-6$ using various sensitivity parameters and calculate relative bias (that is, $(\hat\beta - \beta)/|\beta|\times 100\%$) at each $[\delta_{z_2}, \delta_{y}]$. In addition, if the relative biases are close to $0\%$, we evaluate coverage probabilities of $95\%$ confidence intervals and see if they achieve nominal coverage. Details of IPW and imputation models for submechanisms in the blended analysis are provided in Section \ref{Numerical_Studies1}.

\subsection{Scenario 1: MNAR on covariate}\label{mnar_r22}

\begin{figure}
\centering
\includegraphics[width = 16cm, height = 11cm]{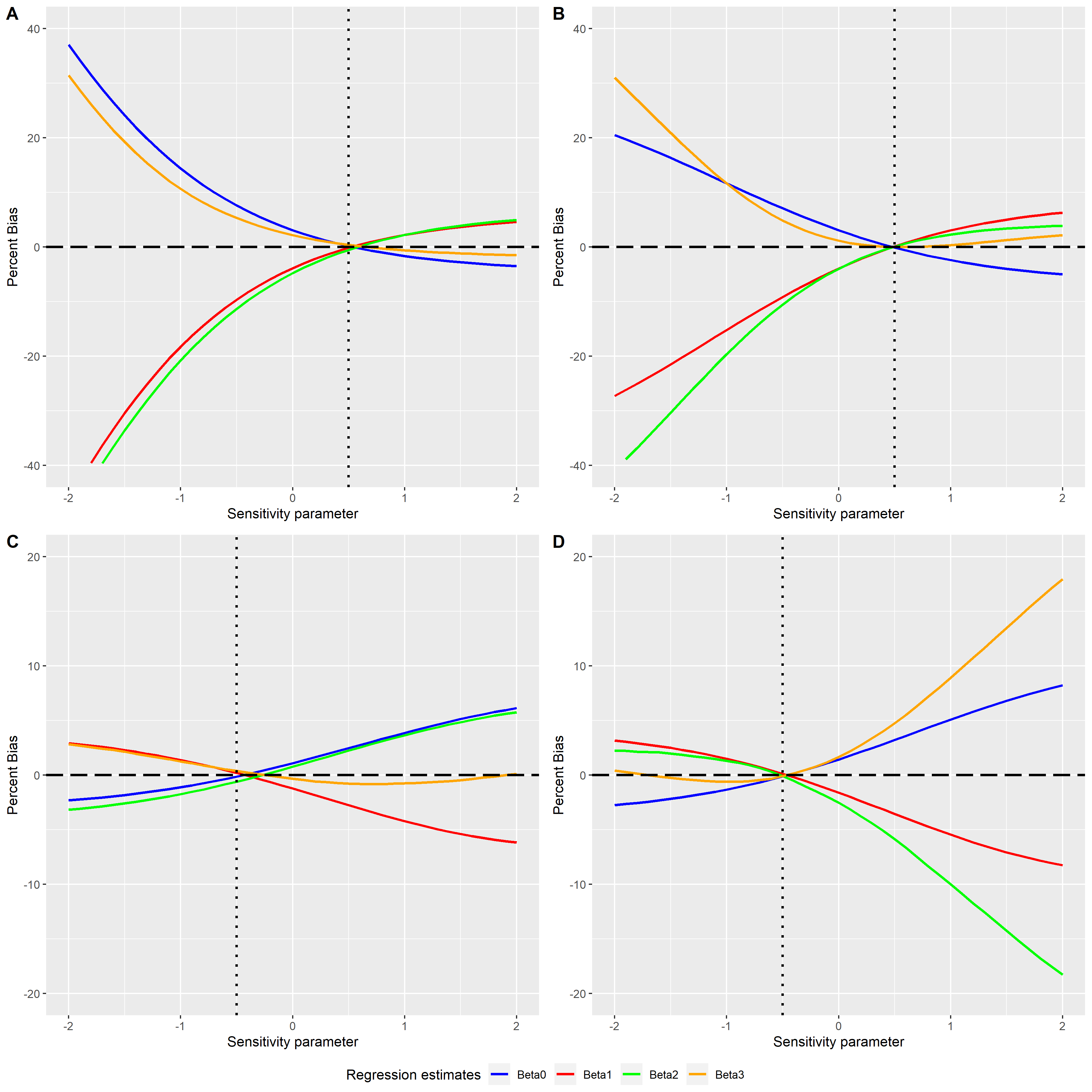}
\caption{Simulation results of Scenario 1. In each plot, the y-axis indicates the percent biases of four regression coefficient estimates, and the x-axis indicates the sensitivity parameter value $\delta_{z_2}$, ranging from $-2.0$ to $2.0$}\label{simulplot_R2}
\end{figure}

In this simulation, we fix $\delta_{y} = 0$ and vary $\delta_{z_2}$ from $-2.0\leq \delta_{z_2}\leq 2.0$ in $0.1$ increments. At each values of $[\delta_{z_2}, \delta_{y}]$, we estimate regression coefficients and evaluate their percent biases. Simulation results are shown in Figure \ref{simulplot_R2}, with one panel for each blended analysis approach. Each individual panel demonstrates how the relative bias changes as a function of $\delta_2$ across the range considered for each of the 4 analysis model parameters $\boldsymbol\beta_{\delta_{z_2}} = [\beta_0, \beta_1, \beta_2, \beta_3]_{\delta_{z_2}}$.  

The top two panels concern ``III" and ``IIM", where missingness in $Z_2$ are addressed via inverse probability weighting. Bias in estimation of all regression coefficient estimates are close to $0\%$ at $\delta_{z_2} = 0.5$, which is  expected because both the sensitivity function and parameter are correctly specified. Similarly, the bottom two panels show relative bias in as a function of $\delta_2$ for ``IMI" and ``IMM", where missingness in $Z_2$ is addressed via multiple imputations. Bias is eliminated at $\delta_{z_2} = -0.4$ for ``IMI" and at $\delta_{z_2} = -0.6$ for ``IMM". We denote $\delta_{Z_2}^*$ as the debiasing sensitivity parameter value for each blended analysis. For all 4 blended analysis procedures, the Bootstrap-MI approach resulted in reasonable coverage for all covariates at $\delta_{z_2}^*$. 

As $\delta_{z_2}$ deviates from $\delta_{Z_2}^*$, the relative bias for all regression coefficients diverges from $0$. Generally, bias was greater when $\delta_{z_2} < \delta^*$ than $\delta_{z_2} \geq \delta^*$ when ``III" and ``IIM" are used, while the opposite trend was observed under ``IMI" and ``IMM". Dramatic bias observed when $\delta_{z_2} < \delta_{z_2}^*$ for  ``III" and ``IIM" is due to the fact that selection probability $P(R_2 = 1\mid X, Z_1^{(l)})$ decreases toward $0$, leading to very large weights $w_2^{(l)}$. 

When a sub-mechanism is addressed via MI, it is not obvious \textit{a priori} which value of $\delta_{z_2}$ will eliminate the bias. This is because the true missingness mechanism is based on a selection model, while imputing under MNAR utilizes a pattern-mixture model-based approach which specifies how the distribution of $Z_2$ differs when it is observed compared to when it is missing. Still, we observe that bias is resolved when $\delta_{z_2} = -0.4$ under ``IMI" and $\delta_{z_2} = -0.6$ under ``IMM". This corresponds to the nature of the missingness mechanism, as $Z_2$ being more likely to be missing when it is 0 than 1 implies that missing values of $Z_2$ are more likely to be 0 than 1, and so the probability of imputing $Z_2$ as $1$ should be adjusted downward.

\subsection{Scenario 2: MNAR on outcome}\label{mnar_r3}

\begin{figure}
\centering
\includegraphics[width = 16cm, height = 11cm]{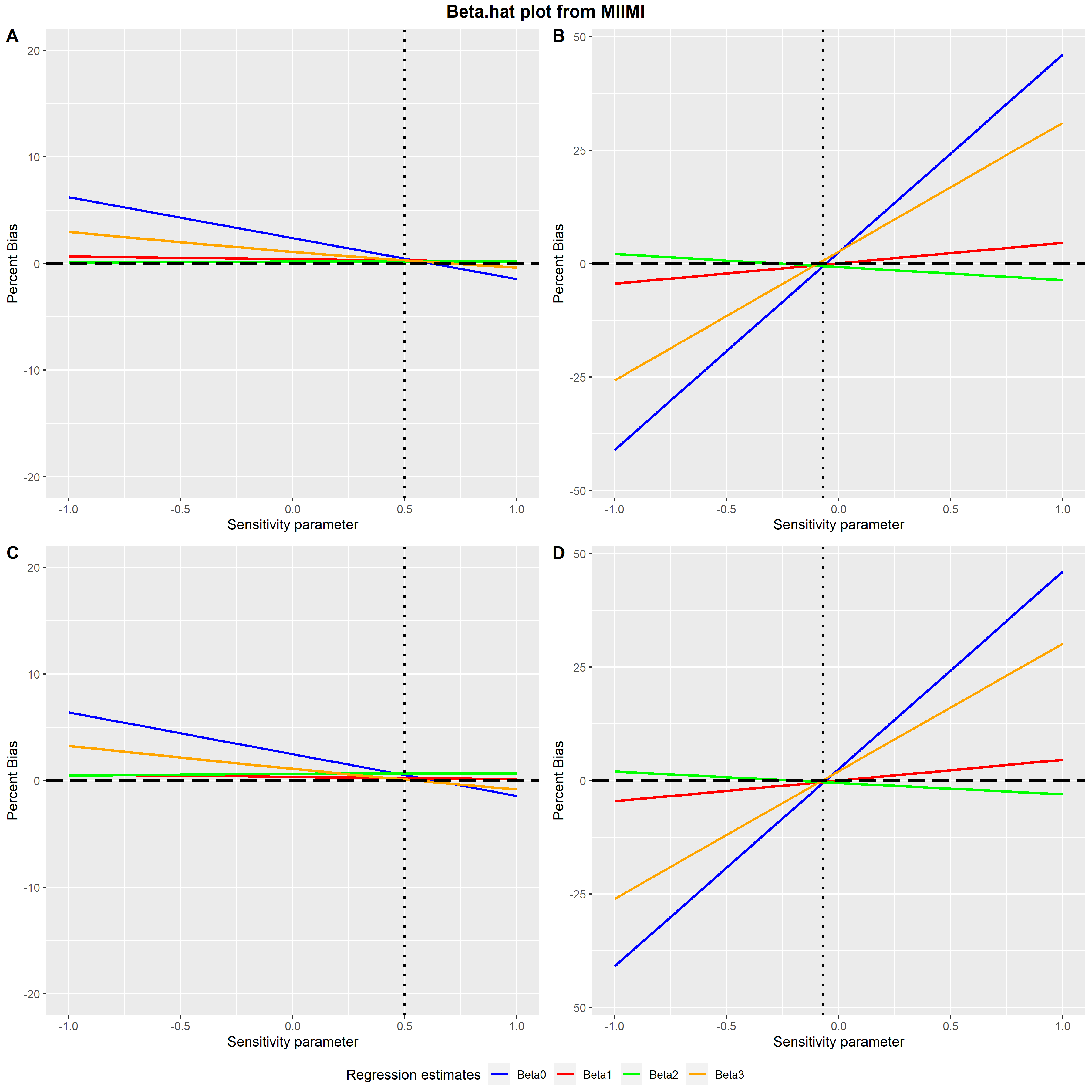}
\caption{Simulation results of Scenario 2. In each plot, the y-axis indicates the percent biases of four regression coefficient estimates, and the x-axis indicates the sensitivity parameter value $\delta_y$, ranging from $-1.0$ to $1.0$}\label{simulplot_R3}
\end{figure}

In this scenario, we fix $\delta_{z_2} = 0$ and vary $\delta_{y}$ from $-1.0\leq\delta_{y}\leq 1.0$ in increments of $0.1$. At each values of $[\delta_{z_2}, \delta_{y}]$, we estimate regression coefficients and evaluate their percent biases. Results are shown in Figure \ref{simulplot_R3}. In ``III" and ``IMI", missingness in $Y$ is addressed via inverse probability weighting (left two panels). We observe that the percent biases of regression estimates are almost completely eliminated when $\delta_{y} = 0.5$. Such a phenomenon is as anticipated, because $\delta_{y} = 0.5$ reflects the true missing data mechanism of outcome variable.

Similarly, the right two panels correspond to ``IIM" and ``IMM", where missingness in $Y$ is addressed via MI. Biases was almost completely removed at $\delta_{y} = -0.08$ for both approaches. Similar to Scenario 1, the particular value of $\delta_y$ that resolve the bias is not immediately obvious from the data generating mechanism and may differ if an alternative sensitivity function is used. Still, it correctly reflects that a smaller values of $Y$ are more likely to be missing, and thus, it is reasonable that imputing smaller values of $Y$ than you would based on the observed data would resolve the bias. Finally, the coverage probabilities of the $95\%$ Bootstrap-MI confidence intervals are also close to $0.95$ in all cases with appropriate $\delta_{y}$. 

\section{Application}\label{application}

A natural challenge in real data applications is that the true data distribution and missing data mechanism are unknown, and thus, the analyst should introduce several assumptions. While the simulation studies in Section \ref{Numerical_Studies} focus on the successful estimation of analysis model parameters at a certain sensitivity function and parameters, the real data application aims to illustrate a set of plausible assumptions and provide details of implementing our proposed method. In such a way, we aim to guide prospective analysts on decision-making in sensitivity analysis under a blended analysis framework, such as determining appropriate ranges of sensitivity parameters and interpreting their implications on analysis results.

\subsection{A study on the weight loss by different surgical type}\label{DURABLE_data}

We employ a subset of the DURABLE study to demonstrate our proposed sensitivity analysis framework in real-data practice. The goal of this section is to illustrate sensitivity analysis to assess violations of the MAR assumption while answering two questions: we focus on patients who received either RYGB or VSG surgical types between 2008 and 2010 in three sites: Kaiser Permanente Northern California, Southern California, and Washington. To mimic the ideal population, we also exclude patients who do not have a BMI greater than 35 in the 365 days leading up to surgery \citep{arterburn2018comparative}. Consequently, we exclude $132$ patients from $9,410$ cohorts and focus on the remaining $9,278$ patients \citep{thaweethai2020statistical}. Table \ref{outcome} illustrates summary statistics of BMIs stratified by surgical types and descriptions of predictors, respectively.

\begin{table}
\centering
\caption{Summary statistics of BMIs by surgical types and descriptions of predictors}\label{outcome}
\begin{tabular}{l|cccc} \hline
Variable & \multicolumn{2}{c}{$BMI_0$} & \multicolumn{2}{c}{$BMI_5$} \\
Surgical type & RYGB & VGA & RYGB & VGA \\ \hline
Mean & $44.85$ & $44.25$ & $34.10$ & $36.42$\\
S.D. & $7.267$ & $7.178$ & $6.781$ & $7.397$\\ \hline
\end{tabular}
\end{table}

As discussed in Section \ref{modularization}, we introduce five sub-mechanisms to modularize the data provenance related to the analysis model of interest. Recall that the analysis model is a multiple linear model with the main effects of $RYGB$ and $CKD_0$ and their interaction, adjusting for other predictors. Table \ref{submechanisms_durable} displays the descriptions of five sub-mechanisms and related sample sizes. 

\begin{table}
\centering
\caption{Five sub-mechanisms, sample sizes, and cumulative/marginal proportions}\label{submechanisms_durable}
\begin{tabular}{c|c|cc} \hline
Sub-mechanisms & Description & Sample size ($R = 1$) & marginal \\\hline
$R_1$ & $1$ if both $CCS_0$ and $CKD_0$ are observed and 0 if not. & $6,700$ & $72.2\%$ \\
$R_2$ & $1$ if $BMI_0$ is observed and 0 if not. & $6,165$ & $89.6\%$ \\
$R_3$ & $1$ if a patient is available beyond 4.5 years after surgery and $0$ if not. & $4,709$ & $75.5\%$ \\
$R_4$ & $1$ if both $CCS_{5}$ and $CKD_{5}$ are observed and $0$ if not. & $2,083$ & $34.3\%$ \\
$R_5$ & $1$ if $BMI_{5}$ is observed and $0$ if not. & $1,922$ & $68.2\%$ \\ \hline 
\end{tabular}
\end{table}

In our analysis, $CKD_0$ is defined based on the ICD-9 code as well as other measurement histories. Specifically, $CKD_0$ has value $1$ if a patient has an ICD-9 code indicating CKD at any point 0-6 months before surgery. $CKD_0$ has value $0$ if a patient (i) does not have ICD-9 indication of CKD and (ii) has some BMI or comorbidity measure during each of the three time periods: 0-2, 2-4, and 4-6 months before surgery. If a patient does not have an ICD-9 indication nor BMI/comorbidity measurement history, we consider $CKD_0$ is missing. Similarly, $CCS_0$ is considered to be observed if a patient's BMI or any comorbid conditions are recorded in the three time windows before surgery and missing otherwise. Lastly, we categorize observed comorbidity scores into three categories: positive, zero, and negative. Follow-up CKD and CCS are similarly defined using a 4.5-5.5 year time window after the surgery. 

In addition, we define $RYGB$ as $1$ if a patient received Roux-en-Y gastric bypass and $0$ otherwise. We also include site (Northern California/Washington/Southern California), gender (male/female), age at the surgery in years, surgery dates in year (2008/2009/2010), race/ethnicity collapsed into four categories (White/Hispanic/Black/Other), and the Charlson-Elixhauser combined comorbidity score ($CCS_0$) with three categories (positive/zero/negative) \citep{Gagne2011} as adjusting factors. Hereafter, we define a vector of covariates ${\bf X} = [RYGB, CKD_0, Site, Gender, Age, Year, Race, CCS_0]$, and specifically, denote ${\bf X}^c$ as a vector of complete covariates: surgical type, Year, Age, Male, and Race. Table \ref{predictors} shows the list and summary statistics of predictor variables. 

\begin{table}
\centering
\caption{Summary statistics of BMIs by surgical types and descriptions of predictors}\label{predictors}
\begin{tabular}{l|c|cc} \hline
Variable & Type & Description \\ \hline
$RYGB$ & Binary & $1$ if surgical type is RYGB ($76.6\%$) and $0$ if VSG ($23.4\%$).\\
$CKD_0$ & Binary & $1$ if one has chronic kidney disease, ($5.01\%$), $0$ if not ($68.1\%$), or missing ($26.9\%$).\\
$Site$ & Categorical & Northern California ($31.5\%$), Washington ($6.88\%$), or Southern California ($61.8\%$).\\
$Male$ & Binary & $1$ if male ($18.3\%$) or not ($81.7\%$).\\
$Age$ & Continuous & Age at the surgery, $29\sim 79$.\\
$Year$ & Categorical & 2008 ($27.2\%$, baseline), 2009 ($30.9\%$), and 2010 ($41.7\%$).\\
$Race$ & Categorical & White ($50.4\%$, baseline), hispanic ($29.4\%$), black ($14.9\%$), or others ($5.3\%$).\\
$CCS_0$ & Categorical & Zero ($29.2\%$, baseline), positive ($27.2\%$), negative ($15.6\%$), or missing ($27.7\%$).\\ \hline
\end{tabular}
\end{table}

\subsection{Blended analysis implementation for DURABLE data}\label{Data_modularization}

As discussed in Section \ref{modularization} and Table \ref{submechanisms_durable}, $[R_1, R_4]$ are related to observing the baseline and follow-up $CKD$ and $CCS$, while $[R_2, R_5]$ are related to the baseline and follow-up BMIs. $R_3$ is the only sub-mechanism unrelated to observing variable(s), and thus IPW is the only possible method; thus, we have $\tilde{\mathcal{K}} = \{1,2,4,5\}$ and there are $2^4 = 16$ possible combinations of $[\mathcal{K}_{IPW}, \mathcal{K}_{MI}]$. However, we focus on two scenarios by ignoring ``IPW only" and ``MI only", and restricting that the follow-up variables receive the same approach as the baseline variables. For example, if $R_2$ is handled via MI, then $R_5$ is also handled using MI because $R_2$ and $R_5$ are related to observing BMIs. Consequently, two possible scenarios of blended analyses are $\mathcal{K}_{IPW} = \{1,4\}$ (denoted as ``IMIIM") and $\mathcal{K}_{IPW} = \{2,5\}$, which denoted as ``MIIMI". 

We assume that the patient disenrollment mechanism $R_3$ is not affected by an unobserved quantity that would have been observed if a patient is active in the system. Thus, we do not posit a sensitivity parameter on $R_3$. The other four mechanisms, however, can be either MAR or MNAR, and thus, the sensitivity analysis should be implemented. Consequently, we introduce a vector of four sensitivity parameters $\boldsymbol{\delta} = [\delta_1, \delta_2, \delta_4, \delta_5]$, where $\delta_j$ corresponds to $R_j$ and $\delta_3 = 0$ all for scenarios. Hereafter, we denote ${\bf X}^{(l)} = [CKD_0^{(l)}, CCS_0^{(l)}, BMI_0^{(l)}, CKD_5^{(l)}, CCS_5^{(l)}, BMI_5^{(l)}]$ be the observed and imputed variables from $l$th imputation. Also, for each k, we define $[\boldsymbol{\hat\alpha}_c, \boldsymbol{\hat\alpha}_{(l)}]$ as the estimated coefficients corresponding to $[{\bf X}_c, {\bf X}^{(l)}]$ in a selection model for $R_k$ and an imputation model for $D_k^m$. Given prespecified sensitivity parameter vectors, the blended analysis for the study sample can be implemented as follows:

\begin{enumerate}
\item Prespecify $\boldsymbol{\delta} = [\delta_1, \delta_2, \delta_4, \delta_5]$.
\item Under ``IMIIM", we fit the selection model for $R_1$ and calculate weight $w_1$ for patients with $R_1 = 1$ as follows:
\begin{align}\label{durable_IPW1}
w_{1}^{-1} &= P(R_1 = 1\mid {\bf X}^c, {\bf X}^{(l)}, CKD_0) = expit(X^c\boldsymbol{\hat\alpha}_c + X^{(l)}\boldsymbol{\hat\alpha}_{(l)} + \delta_{1}CKD_0).
\end{align}
Under ``MIIMI", we estimate the imputation model for $[CKD_0, CCS_0]$ using patients with $R_{i1} = 1$ as follows:
\begin{align}\label{durable_MI1}
P(CKD_0 = 1\mid {\bf X}^c, {\bf X}^{(l)}) &= expit({\bf X}^c\boldsymbol{\hat\alpha}_c + X^{(l)}\boldsymbol{\hat\alpha}_{(l)} + \delta_1(1 - R_1)), \nonumber\\
P(CCS_0 = k\mid {\bf X}^c, {\bf X}^{(l)}) &= expit({\bf X}^c\boldsymbol{\hat\alpha}_{k,c} + X^{(l)}\boldsymbol{\hat\alpha}_{k,(l)} + \delta_1(1 - R_1)),~k=1,2,
\end{align}
where $CCS_0 = 1$ implies ``postive" and $CCS_0 = 2$ implies ``negative". Using the imputation model, we draw $CKD_{i0}^{m(l)}$ and $CCS_{i0}^{m(l)}$ for a patient $i$ with $R_{i1} = 0$.
\item Under ``IMIIM", we fit the selection model for $R_2$ and calculate weight $w_2$ for patients with $R_2 = 1$ as follows:
\begin{align}\label{durable_IPW2}
w_{2}^{-1} & = P(R_2 = 1\mid {\bf X}^c, {\bf X}^{(l)}, BMI_0, \bar{R}_1 = 1) = expit({\bf X}^c\boldsymbol{\hat\alpha}_c + {\bf X}^{(l)}\boldsymbol{\hat\alpha}_{(l)} + \delta_{2}BMI_0/5).
\end{align}
Under ``IMIIM", we estimate the imputation model for $BMI_{0}^{m}$ using patients with $R_{i2} = 1$ as follows:
\begin{align}\label{durable_MI2}
E(BMI_0\mid {\bf X}^c, {\bf X}^{(l)}) &= {\bf X}^c\boldsymbol{\hat\alpha}_c + {\bf X}^{(l)}\boldsymbol{\hat\alpha}_{(l)} + \delta_2(1 - R_2).
\end{align}
Using the imputation model, we draw a random sample of $BMI_{i0}^{m(l)}$ for a patient $i$ with $R_{i2} = 0$.

\item For $R_3$, let $T$ be the time to disenrollment. We fit a Cox proportional hazard model and estimate the baseline survival function $S_0(t)$ using the Breslow estimator \citep{Breslow1972} as follows:
\begin{align}\label{durable_IPW3}
\lambda(T) &= \lambda_0(T)exp({\bf X}^c\boldsymbol{\hat\alpha}_c + {\bf X}^{(l)}\boldsymbol{\hat\alpha}_{(l)}).
\end{align}
The inverse probability weight $w_3$ is calculated as follows:
\begin{align}
w_3^{-1} &= P(R_3 = 1\mid {\bf X}^c, {\bf X}^{(l)}) = \hat{S}_0(4.5)^{\exp({\bf X}^c\boldsymbol{\hat\alpha}_c + {\bf X}^{(l)}\boldsymbol{\hat\alpha}_{(l)})}.
\end{align}
\item Under ``IMIIM", we fit the selection model for $R_4$ and calculate weight $w_4$ for patients with $R_4 = 1$ as follows:
\begin{align}\label{durable_IPW4}
w_{4}^{-1} &= P(R_4 = 1\mid \bar{R}_3 = 1, {\bf X}_c, CKD_5, CKD_0^{m(l)}, CCS_0^{m(l)}, BMI_0^{m(l)}) = expit({\bf X}^c\boldsymbol{\hat\alpha}_c + {\bf X}^{(l)}\boldsymbol{\hat\alpha}_{(l)} + \delta_{4}CKD_5).
\end{align}
Under ``MIIMI", we estimate the imputation model for $[CKD_{5}^{m}, CCS_{5}^{m}]$ using patients with $R_{i4} = 1$ as follows:
\begin{align}\label{durable_MI4}
P(CKD_5 = 1\mid {\bf X}^c, {\bf X}^{(l)}) &= expit({\bf X}^c\boldsymbol{\hat\alpha}_c + {\bf X}^{(l)}\boldsymbol{\hat\alpha}_{(l)} + \delta_4(1 - R_1)), \nonumber\\
P(CCS_5 = k\mid {\bf X}^c, {\bf X}^{(l)}) &= expit({\bf X}^c\boldsymbol{\hat\alpha}_{k,c} + {\bf X}^{(l)}\boldsymbol{\hat\alpha}_{k,(l)} + \delta_4(1 - R_1)),~k=1,2,
\end{align}
where $CCS_5 = 1$ implies ``postive" and $CCS_5 = 2$ implies ``negative". Using the imputation model, we draw random samples of $CKD_{i5}^{m(l)}$ and $CCS_{i5}^{m(l)}$ for a patient $i$ with $R_{i4} = 0$.

\item Under ``MIIMI", we fit the selection model for $R_5$ and calculate weight $w_5$ for patients with $R_5 = 1$ as follows:
\begin{align}\label{durable_IPW5}
w_{5}^{-1} & = P(R_5 = 1\mid {\bf X}_c, {\bf X}^{(l)}, \bar{R}_4 = 1) = expit({\bf X}^c\boldsymbol{\hat\alpha}_c + {\bf X}^{(l)}\boldsymbol{\hat\alpha}_{(l)}+ \delta_{5}BMI_0/5).
\end{align}
Under ``IMIIM", we estimate the imputation model for $BMI_{5}^{m}$ using patients with $R_{i5} = 1$ as follows:
\begin{align}\label{durable_MI5}
E(BMI_5\mid {\bf X}^c, {\bf X}^{(l)}) &= {\bf X}^c\boldsymbol{\hat\alpha}_c + {\bf X}^{(l)}\boldsymbol{\hat\alpha}_{(l)} + \delta_5(1 - R_5).
\end{align}
Using the imputation model, we draw a random sample of $BMI_{i5}^{m(l)}$ for a patient $i$ with $R_{i5} = 0$.
\item Fit the analysis model using observed/imputed data set $[{\bf X}^c, {\bf X}^{(l)}]$ and obtain $\boldsymbol{\beta}_{\boldsymbol\delta}^{(l)}$. Repeat $2.\sim 6.$ $10$ times and obtain $\boldsymbol{\beta}_{\boldsymbol\delta} = \frac{1}{10}\sum\limits_{l=1}^{10}\boldsymbol{\beta}_{\boldsymbol\delta}^{(l)}$.
\end{enumerate}

As introduced in Sections \ref{MI} and \ref{IPW}, a sensitivity parameter $\delta_k$ plays different roles depending on whether IPW or MI is used for each sub-mechanism $R_k$. When an IPW is used, $exp(\delta_k)$ represents the multiplicative increment of the odds ratio of $R_k = 1$ and $R_k = 0$. Under MI, $\delta_k$ represents the additive increment of average imputed values (for continuous outcome) or average probability in the logit scale. For example, $\delta_2$ in Eq. (\ref{durable_IPW2}) implies that the odds of $R_2 = 1$ (versus $R_2 = 0$) increases by $exp(\delta_2)$ as $BMI_0$ increases by $5$ unit. In Eq. (\ref{durable_MI2}), $\delta_2$ represents the additive increase in the average imputed $BMI_0$ values. 

\subsection{Strategies for choosing ranges of sensitivity parameters}\label{Sec5_delta_interval}

One of the main concerns about conducting a sensitivity analysis is choosing plausible sensitivity parameter values. Several approaches and sources of information can be used for this purpose. One approach is to reflect subject-specific expertise beliefs on the missing data mechanism \citep{white2007eliciting}. These experts can be anyone who can contribute knowledge in understanding the missing data, such as trial investigators, clinicians, or patients \citep{mason2017development}. The expert-driven suggestions can be reflected in the analysis via reasonable ranges for the sensitivity parameters and successfully capture the possible fluctuations of analysis results or potentially rule out the concerns on MNAR missing values (if the analysis results do not show noticeable changes over varying sensitivity parameters).

Another approach is to implement a ``tipping point" analysis by changing the sensitivity parameter until a different conclusion is reached (for example, being or not being significant). The analyst can then discuss the plausibility of this value with the subject-specific experts. This approach is appealing because it can easily be implemented and requires less effort in decision-making than formal elicitation. On the other hand, such ``tipping point" analysis depends on the definition of ``conclusion," which may be subjected to an analysis model and can be challenging to implement with multiple sensitivity parameters. 

Although sensitivity analysis in the blended analysis framework offers flexibility via multiple sub-mechanisms, the contribution of a single sensitivity parameter is hard to understand if multiple sensitivity parameters move simultaneously. Consequently, the first step is to implement a series of conditional sensitivity analyses over selected intervals for each sub-mechanism while holding other sub-mechanisms to be MAR. As two or more sub-mechanisms present highly noticeable fluctuation under given sensitivity parameter ranges, we proceed to multi-dimensional sensitivity analysis by considering multi-dimensional surfaces of analysis results.

One strategy of choosing the ranges of each sensitivity parameter is to based on its implication on the corresponding selection probability or imputation model. When a sub-mechanism $R_k$ is addressed via MI, the average imputed values of $D_k^m$ can be used to explicitly understand the effect of a particular $\delta_k$ on the analysis. For example, if the blended analysis is implemented under ``IMIIM", $\delta_2$ and $\delta_5$ represent the additive increase in average imputed BMI at baseline and follow-up, respectively. The average imputed baseline BMI at $\delta_2 = 0$ is $46.178$, and moves between $[40.178, 52.178]$ as $\delta_2$ moves between $[-6.0, 6.0]$. Based on such implications of a particular $\delta_2$ on unobserved baseline BMI, the analyst may specify appropriate ranges of sensitivity parameters to properly reflect the MNAR missing data mechanism of concern.

When IPW is used for $R_k$, implications of a particular $\delta_k$ on the blended analysis result can be explicitly evaluated via the distribution of $D_k^m$. The distribution of $D_k^m$ given $\delta_k$ can be summarized using an appropriate ``connecting quantity". For example, when $R_1$ is modeled via IPW, the conditional probability $P(CKD_0 = 1 \mid R_1 = 0, \delta_1)$ reflects the implication of $\delta_1$ on the distribution of unobserved baseline CKD. Similarly, when the ``MIIMI" option is used for the blended analysis, $R_2$ is modeled via IPW. In this case, a quantity of interest is $E(BMI_0\mid R_2 = 0, \delta_2)$, the average of the unobserved baseline BMI. The conditional probabilities and expectations of unobserved groups are unobservable, but we can estimate these quantities using the conditional probabilities specified in the blended analysis. Details of calculations of connecting quantities are provided in Section \ref{Sec5_delta_interval}. Note that such connecting quantities are subject to the sensitivity parameter and thus provide the implications of a particular sensitivity parameter value on the unobserved data. In such a way, behaviors of these quantities can be used to choose appropriate ranges of sensitivity parameters, which also requires considering expert knowledge of domains.

\subsection{Analysis results}\label{Sec5_analysis_results}

As an initial analysis, we implement the blended analysis as elaborated in Section \ref{Data_modularization} under MAR assumption by holding all sensitivity parameters as $0$. Table \ref{Durable_MAR} illustrates the estimated main effects of surgical type, CKD indication at baseline, and their interaction using ``IMIIM" and ``MIIMI". Analysis result of the blended analysis using ``IMIIM" is based on $2,262$ patients with $R_{1} = R_{3} = R_{4} = 1$. As shown in Table \ref{Durable_MAR}, the average PTBC is $2.51$ higher in a patient with baseline CKD received RYGB and $5.93$ higher if a patient does not have CKD at baseline, adjusting for other predictors. Similarly, the blended analysis using ``MIIMI" is based on $5,115$ patients with $R_{2} = R_{3} = R_{5}= 1$. The average PTBC is $4.82$ higher in a patient with baseline CKD received RYGB and $6.79$ higher if a patient does not have CKD at baseline, conditioning on other predictors. Lastly, $95\%$ confidence intervals are obtained via the bootstrap-MI procedure.

\begin{table}
\centering
\caption{The estimated regression coefficients of surgical type, chronic kidney disease indication at baseline, and their interaction using blended analyses (``IMIIM" and ``MIIMI")}\label{Durable_MAR}
\begin{tabular}{c|cc} \hline 
& IMIIM & MIIMI \\\hline
Intercept & -21.7 [-24.8, -18.6] & -21.2 [-23.1, -19.3] \\
$RYGB$ & -5.9 [-7.3, -4.6] & -6.8 [-7.6, -6.0] \\
$CKD_0$ & -4.8 [-9.1, -0.5] & -2.5 [-5.2, 0.2] \\
$RYGB\times CKD_0$ & 3.4 [-1.4, 8.2] & 2.0 [-0.98, 5.0] \\\hline
\end{tabular}
\end{table}

\subsubsection{Conditional sensitivity analysis}\label{Conditional_sensitivity_analysis}

Using the results under MAR as the origin, we investigate the impact of the missing data mechanism for each sub-mechanism departing from MAR toward MNAR by repeating the analysis using various values of $[\delta_1,\delta_2,\delta_4,\delta_5]$. For all $k\in \{1,2,4,5\}$, the range of $\delta_k$ is $[-2.0, 2.0]$ if $k\in \mathcal{K}_{IPW}$ and $[-6.0, 6.0]$ if $k\in \mathcal{K}_{MI}$. To implement the sensitivity analysis of analysis results, we repeat the blended analysis over all combinations of $[\delta_1, \delta_2, \delta_4, \delta_5]$, where each $\delta_k$ for IPW moves by $0.1$ unit, and $\delta_k$ for MI moves by $0.3$ unit. For example, when implementing ``IMIIM", values of $\delta_1$ and $\delta_4$ are $[-2.0, -1.9,\ldots, 2.0]$, while $\delta_2$ and $\delta_5$ values are $[-6.0, -5.7,\ldots, 5.7, 6.0]$. Such $\boldsymbol{\delta}$ values yield $41^4 = 2,825,761$ sets of regression estimates.

\begin{figure}
\centering
\includegraphics[width = 18cm, height = 13cm]{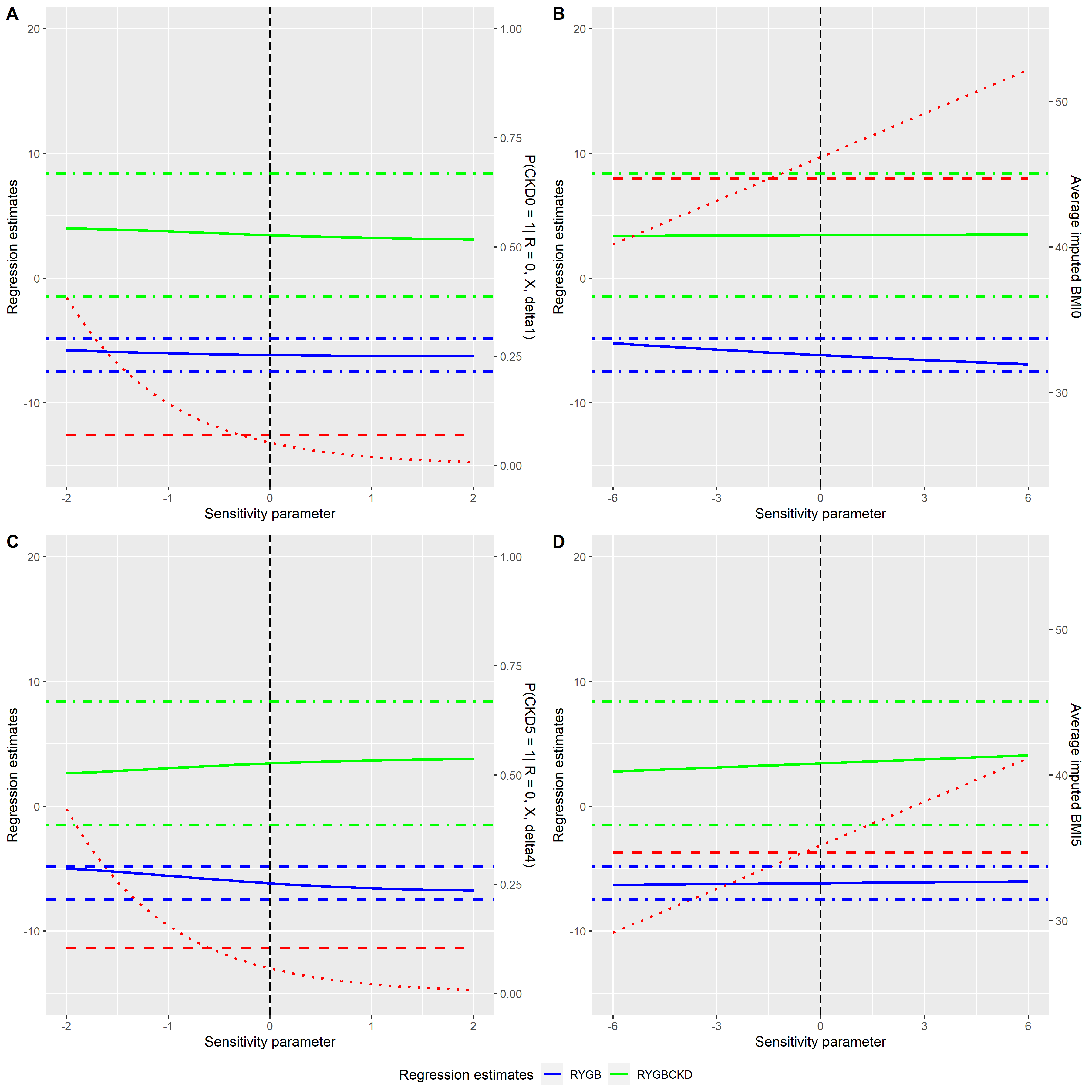}
\caption{Conditional sensitivity analysis using IMIIM. The left axis in each plot presents regression estimates of surgical type and its interaction with baseline CKD, and their $95\%$ CI evaluated under MAR assumption. Right axes in plots A and C describe the proportion of unobserved baseline CKD (plot A) and follow-up CKD (plot C), in which the red dotted curves correspond to. Right axes in plots B and D describe the average imputed values of baseline BMI (plot B) and follow-up BMI (plot D), in which the red dotted curves correspond to. The red horizontal dashed lines indicate the observed proportion of CKD (plots A and C) and average of BMI (plots B and D).}\label{IMIIM}
\end{figure}

The first step is conditional sensitivity analyses, where we vary a single $\delta_k$ while holding others as $0$. Figure \ref{IMIIM} presents conditional sensitivity analysis results using ``IMIIM", where $R_2$ and $R_5$ are addressed via MI, and others are addressed using IPW. In Figure \ref{IMIIM}, each plot presents fluctuations in point estimates of surgical type ($\hat\beta_{RYGB}$, presented in blue solid lines) and its interaction with baseline CKD ($\hat\beta_{RYGB\times CKD_0}$, presented in green, solid lines) over $\delta_k$, and their $95\%$ CI evaluated at $\delta_k = 0$ (presented in dot-dashed lines). The left y-axis represents the magnitudes of the estimates and their CIs, while the right y-axis represents either the average selection probability (i.e., the average of $P(\bar{R}_{i5} = 1)$ across all subjects with $\bar{R}_5 = 1$), or the average imputed values of $BMI_0$ or $BMI_5$ over $\delta_k$. Lastly, the red-colored dashed horizontal lines indicate the averages of observed baseline/follow-up BMI and proportions of observed patients with baseline/follow-up CKD indication.

Plot A in Figure \ref{IMIIM} presents the fluctuations of results over $\delta_1$ conditioning on other $\delta_k$ as $0$. As specified in Eq. (\ref{durable_IPW1}), the odds of $R_1 = 1$ (versus $R_1 = 0$) for patients with CKD at baseline increases by $exp(\delta_1)$ times compared to the odds at $\delta_1 = 0$. Consequently, as $\delta_1$ increases from $-2$ to $2$, the estimated proportion of baseline CKD among unobserved patients decreases from $0.384$ to $0.007$, as indicated in the right y-axis. Similarly, plot C in Figure \ref{IMIIM} presents the fluctuations of results over $\delta_4$ conditioning on other $\delta_k$ as $0$. The odds of $R_4 = 1$ (versus $R_4 = 0$) for patients with CKD at baseline increases by $exp(\delta_4)$ times compared to the odds at $\delta_4 = 0$, and the estimated proportion of follow-up CKD indication among unobserved patients decreases from $0.423$ to $0.008$, as $\delta_4$ increases from $-2$ to $2$.

Next, plot B in Figure \ref{IMIIM} presents results fluctuations over different $\delta_2$, and their $95\%$ CI under $\delta_2 = 0$. The left y-axis represents the magnitudes of the estimates and their CIs, similar to the plots on the left. The right y-axis in this plot represents the average imputed BMI at baseline, the average of imputed $BMI_0$ across all subjects with $R_2 = 0$. As specified in Eq. (\ref{durable_MI2}), $\delta_2$ represents the increment in the average imputed value of $BMI_0$ compared to the average imputed $BMI_0$ under MAR condition. Consequently, as $\delta_2$ increases from $-6$ to $6$, the average imputed $BMI_0$ increases from $40.18$ to $52.18$, as indicated in the right y-axis. Similarly, plot D in Figure \ref{IMIIM} presents results fluctuations over different $\delta_5$, and their $95\%$ CIs under $\delta_5 = 0$. As $\delta_2$ increases from $-6$ to $6$, the average imputed $BMI_0$ increases from $29.18$ to $41.18$, as shown the right y-axis.

\begin{figure}
\centering
\includegraphics[width = 18cm, height = 13cm]{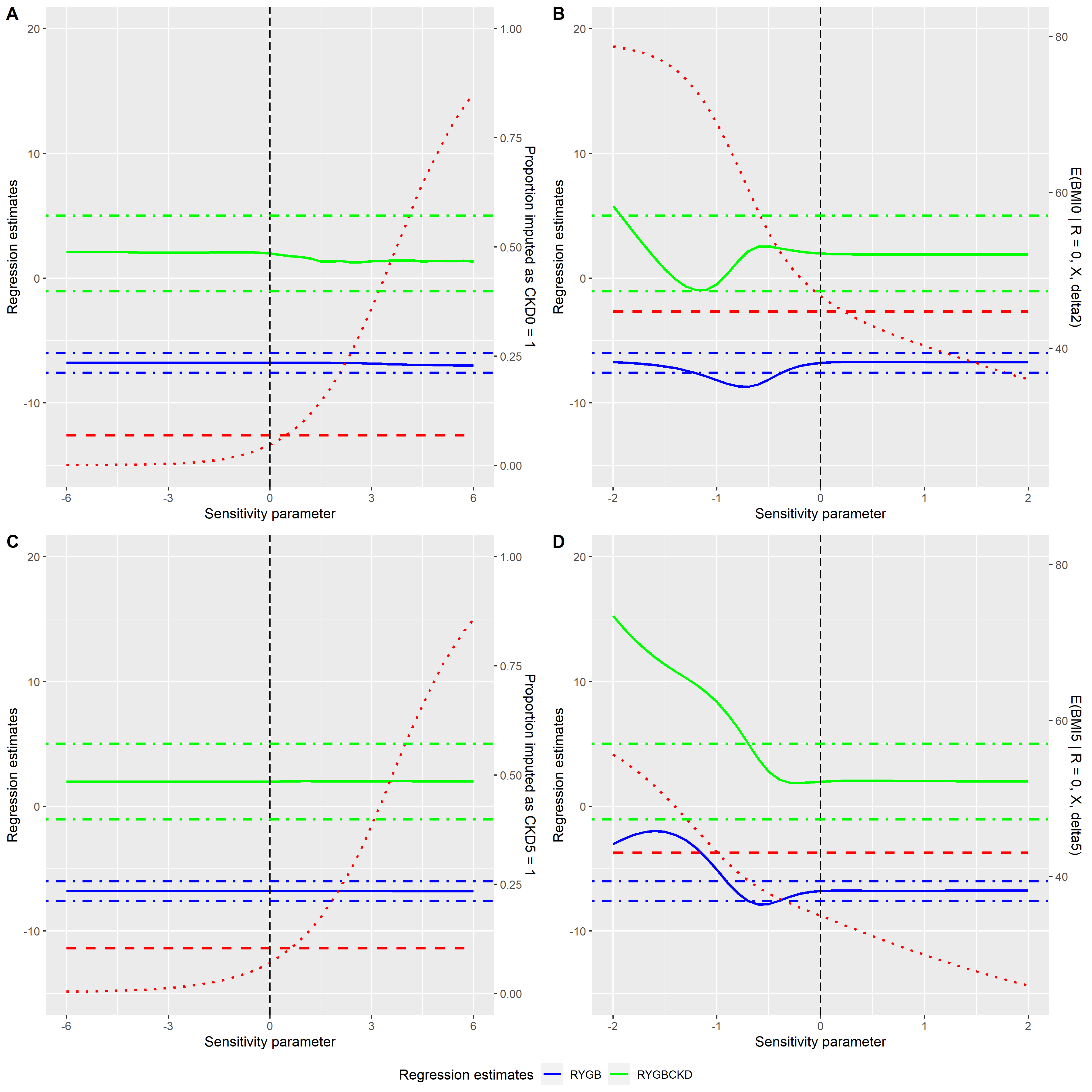}
\caption{Conditional sensitivity analysis using MIIMI. The left axis in each plot presents regression estimates of surgical type and its interaction with baseline CKD, and their $95\%$ CI evaluated under MAR assumption. Right axes in plots A and C describe the average imputed proportion of unobserved baseline CKD (plot A) and follow-up CKD (plot C), in which the red dotted curves correspond to. Right axes in plots B and D describe the average values of unobserved baseline BMI (plot B) and follow-up BMI (plot D), in which the red dotted curves correspond to. The red horizontal dashed lines indicate the observed proportion of CKD (plots A and C) and average of BMI (plots B and D).}\label{MIIMI}
\end{figure}

Figure \ref{MIIMI} presents conditional sensitivity analysis results using ``MIIMI", where $R_1$ and $R_4$ are addressed via MI, and others are addressed using IPW. Similar to Figure \ref{IMIIM}, each plot in Figure \ref{MIIMI} presents point estimates of surgical type (blue solid line) and its interaction between surgical type and baseline CKD (green solid line) over $\delta_k$, and their $95\%$ CIs evaluated at $\delta_k = 0$ (presented in dot-dashed lines). Also, the left and right y-axes represent the magnitudes of the estimates, and either the average selection probability or the average proportions of patients imputed as $CKD_0$ or $CKD_5$ as $1$ over $\delta_k$.

The top-left plot in Figure \ref{MIIMI} presents the fluctuations of regression estimates over $\delta_1$ conditioning on other $\delta_k$ as $0$. As specified in Eq. (\ref{durable_MI1}), the odds of $CKD_{0} = 1$ versus $CKD_{0} = 0$ increase by $exp(\delta_1)$ times compared to the odds at $\delta_1 = 0$. Consequently, the average proportion of patients imputed as $CKD_0 = 1$ is $0.047$ at $\delta_1 = 0$, and as $\delta_1$ increases from $-6$ to $6$, the proportion of baseline CKD among unobserved increases from $0.001$ to $0.853$, as indicated on the right y-axis. Similarly, the bottom-left plot presents the fluctuations of regression estimates over $\delta_4$ conditioning on other $\delta_k$ as $0$. The average proportion of patients imputed as $CKD_5 = 1$ is $0.070$ at $\delta_4 = 0$ it increases from $0.003$ to $0.858$, as $\delta_4$ increases from $-6$ to $6$. 

Finally, the top-right plot in Figure \ref{MIIMI} presents the results fluctuations over $\delta_2$ conditioning on other $\delta_k$ as $0$. As specified in Eq. (\ref{durable_IPW2}), the odds of $R_2 = 1$ (versus $R_2 = 0$) for patients increases by $5exp(\delta_1)$ times as their $BMI_0$ increase by 1 unit. This implies that patients with high $BMI_0$ are more likely to be observed if $\delta_2 > 0$, and vice versa. Specifically, the estimated average unobserved baseline BMI decreases from $78.70$ to $36.02$ as $\delta_2$ increases from $-2$ to $2$. Note that the regression estimates dramatically fluctuate when $\delta_2 < 0$ but barely move when $\delta_2 > 0$. This is because as $\delta_2$ decreases to $-2$, the average selection probabilities reduce toward $0.097$, and thus their inverse (which is, the IPW) inflate to a large number. Similar trends can be found in the bottom-right plot, where the fluctuations of regression estimates over $\delta_5$ are illustrated. As $\delta_5$ increases from $-2$ to $0$, the parameter estimates of the surgical type and its interaction with $CKD_0$ dramatically fluctuate. Such noticeable variations over $\delta_5$ imply that analysis results are sensitive to the MNAR assumption on $R_5$, the missingness of the follow-up BMI.

\subsubsection{Two-dimensional sensitivity analysis}\label{Twoway_sensitivity_analysis}

From the conditional sensitivity analyses on each sub-mechanism of ``MIIMI" and ``IMIIM", we discovered that the regression coefficients of ``MIIMI" are highly sensitive to the choice of $\delta_2$ and $\delta_5$ when other sensitivity parameters are fixed to $0$. As the next step, we implement two-dimensional sensitivity analyses on regression estimates of ``MIIMI" over $[\delta_2, \delta_5]$, where ranges of $\delta_2$ and $\delta_5$ both $[-2, 2]$ and moving by $0.1$. To do this, we collect regression estimates $[\hat\beta_{RYGB}, \hat\beta_{RYGB\times CKD_0}]$ from all combinations of four sensitivity parameters $[\delta_1,\delta_2,\delta_4, \delta_5]$. Next, we investigate two-dimensional heat-maps of $\hat\beta_{RYGB}$ and $\hat\beta_{RYGB\times CKD_0}$ over $\delta_2$ and $\delta_5$, accounting for variations over $[\delta_1, \delta_4]$. In each heat map, parameter estimates are presented via the gradation of colors, so the changes of color illustrate fluctuations of $\hat\beta_{RYGB}$ over $[\delta_2, \delta_5]$.

\begin{figure}
\centering
\includegraphics[width = 14cm, height = 10cm]{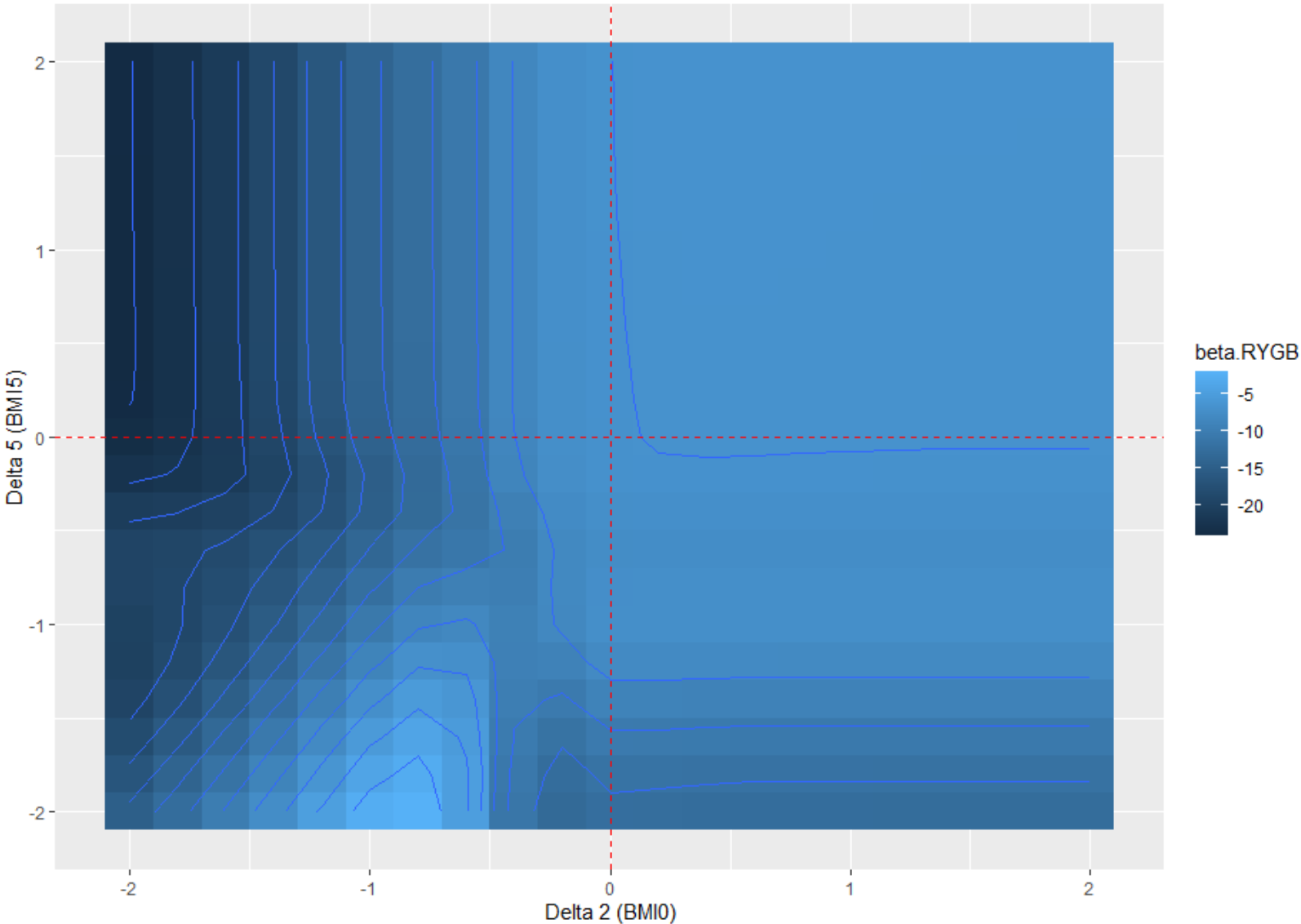}
\caption{Two-way sensitivity analysis on RYGB when MIIMI is used. The x-axis indicates $\delta_2$ values that represent the magnitude of MNAR missingness on $R_2$, while the y-axis indicates $\delta_5$ values that represent the magnitude of MNAR missingness on $R_5$. The gradation of the heatmap indicate the estimated regression coefficient of the surgical type.}\label{RYGB25}
\end{figure}

Figure \ref{RYGB25} illustrates the estimated coefficients of $RYGB$ over different values of $[\delta_2, \delta_5]$. Variations of $\hat\beta_{RYGB}$ (that is, the main effect of surgical types) over $[\delta_2, \delta_5]$ are negligible when both $\delta_2$ and $\delta_5$ are positive. Such trends are consistent with the conditional sensitivity results illustrated in Figure \ref{MIIMI}. Noticeable variations of $\hat\beta_{RYGB}$ are observed elsewhere. For example, $\hat\beta_{RYGB}$ monotonously decreases as one of $\delta$ decreases to $-2$. As both $\delta_2$ and $\delta_5$ decrease to $-2$, variations of $\hat\beta_{RYGB}$ are unpredictable. Such patterns imply that $\hat\beta_{RYGB}$ is robust to the choices of positive values on $\delta_2$ and $\delta_5$ but is highly sensitive otherwise.

\begin{figure}
\centering
\includegraphics[width = 14cm, height = 10cm]{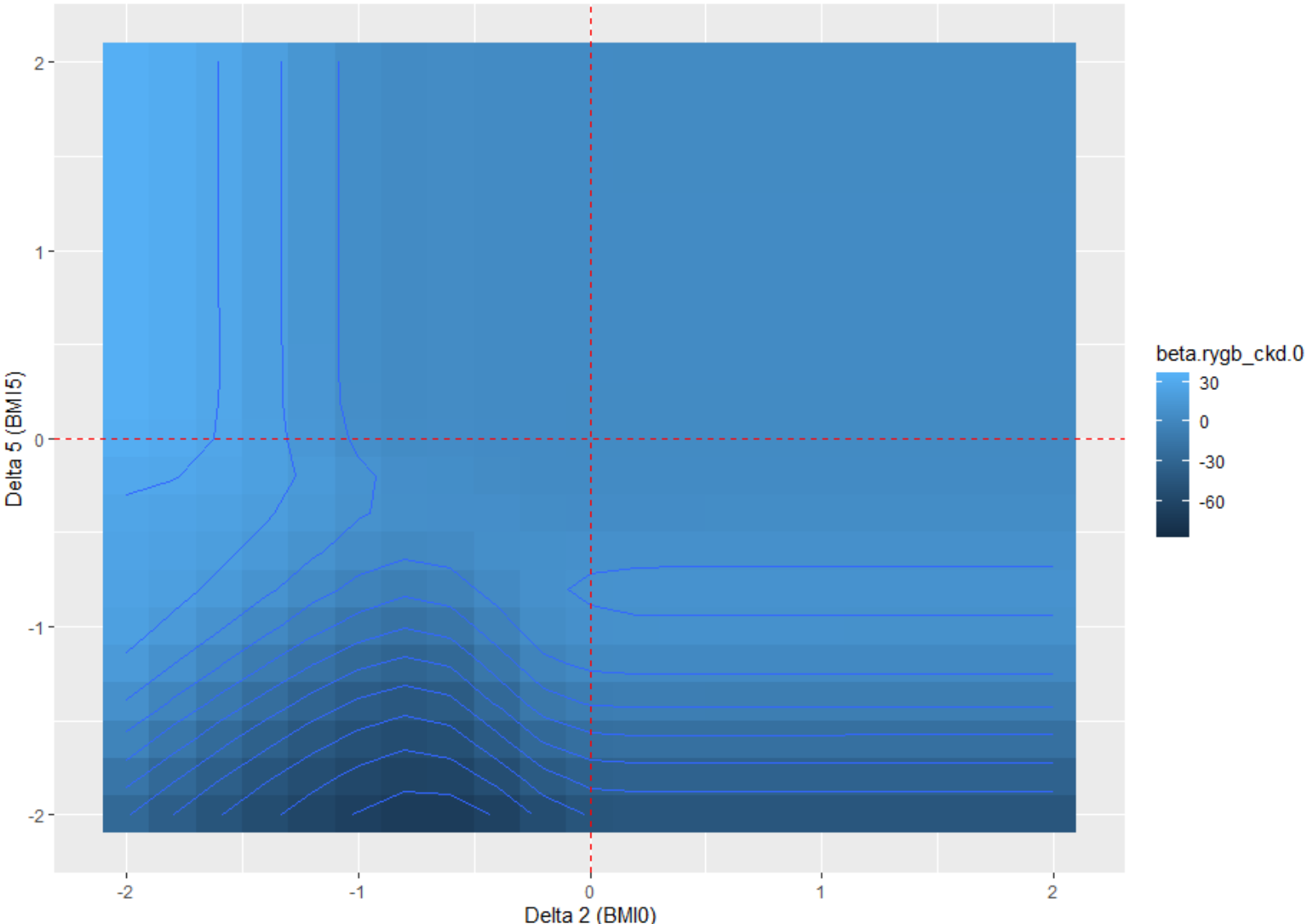}
\caption{Two-way sensitivity analysis on RYGB when MIIMI is used. The x-axis indicates $\delta_2$ values that represent the magnitude of MNAR missingness on $R_2$, while the y-axis indicates $\delta_5$ values that represent the magnitude of MNAR missingness on $R_5$. The gradation of the heatmap indicate the estimated regression coefficient of the interaction terms between surgical type and the baseline $CKD$ indicator.}\label{RYGB_CKD25}
\end{figure}

Finally, Figure \ref{RYGB_CKD25} illustrates the estimates of interaction between $RYGB$ and $CKD$ over different values of $[\delta_2, \delta_5]$ accounting for the variations over $[\delta_1, \delta_4]$. Similar to Figure \ref{RYGB25}, variations of $\hat\beta_{RYGB\times CKD_0}$ over $[\delta_2, \delta_5]$ are negligible when both $\delta_2$ and $\delta_5$ range between $[-0.5, 2.0]$. On the other hand, variations of $\hat\beta_{RYGB\times CKD_0}$ are monotone when $[\delta_2 < -0.5, \delta_5 > 0]$ and $[\delta_2 > 0, \delta_5 < 0]$. Such variations of regression coefficients become more dramatic when both $\delta_2$ and $\delta_5$ are negative, presenting high sensitivity to the choices of $\delta_2$ and $\delta_5$. 

\section{Discussions}\label{Discussions}

The blended analysis framework has been suggested to address missing data problems in EHR-based research. The philosophy of blended analysis is to decompose the data provenance into a sequence of sub-mechanisms representing variables' missing patterns and clinical decisions triggering missing values. Each sub-mechanism is then modeled via IPW or MI based on the appropriate assumptions of its behavior, including missing data mechanisms such as MAR or MNAR. Since the observed data cannot determine whether the MAR assumption is acceptable or not, the sensitivity analysis can be helpful for investigating the robustness of conclusions toward MAR assumptions. In the blended analysis framework, we propose implementing sensitivity analysis by tailoring the selection/imputation model using $\delta$-adjustment method, which can be easily implementable and interpreted. For each sub-mechanism, the selection/imputation model is a function of both observed data and a sensitivity parameter, where the sensitivity parameter represents the linear association between unobserved data and its missingness. By exploring results over different sensitivity parameters, one may investigate how sensitive the analysis results are to the MAR assumption. For example, we conclude that the analysis result is not robust to the MAR assumption if the analysis results dramatically change over sensitivity parameter values. Such a sensitivity analysis may guide understanding the impact of violation of the MAR assumption on the analysis result. 


We implemented several numerical studies to examine the performances of our proposed method. Simulation studies validate the blended analysis may yield unbiased estimations and inferences if both the sensitivity function and parameter are correctly specified. Simulation studies also reveal that the blended analysis results present biases in point estimates when an incorrect sensitivity parameter is specified, indicating that an incorrect sensitivity parameter may mislead the conclusion. Such properties ensure that our proposed framework can be employed for sensitivity analysis and investigate whether a conclusion from a set of particular assumptions is robust to the violations of the assumptions or not. 

We also present an application of blended analysis to real data and delineate the steps to perform sensitivity analyses of MAR assumptions on multiple sub-mechanisms. The purpose of the data application is to demonstrate how the proposed sensitivity analysis can be practically performed when implementing a blended analysis, which requires several decisions on how to establish selection/imputation models for sub-mechanisms. As a real-data example, we employed a subset of the DURABLE data set and delineated steps of blended analysis, investigating the effect of surgical types and their interactions with baseline CKD indication on the five-year BMI change after bariatric surgery. 


Conclusions from sensitivity analyses are subject to the prespecified sensitivity functions and ranges of sensitivity parameters. For a given form of a sensitivity function, ranges of sensitivity parameters can be determined based on knowledge from other studies for similar research goals or experts. For each sub-mechanism, we investigate the conditional expectation of a variable of interest. To streamline the interpretation of a particular sensitivity parameter value and clarify its implication on the distribution of unobserved variables of interest. When MI is used, the average imputed value is used to assess the impact of a sensitivity parameter value on the distribution of unobserved data. Similarly, we calculate the conditional expectation of unobserved variables when IPW is used. Such summary statistics can be used as a communication tool between analysts and experts with domain knowledge and can be used to find a reasonable range of sensitivity parameters.

\bibliographystyle{plainnat}
\bibliography{Referencelist.bib}

\appendix

\section{Details of numerical studies}\label{Numerical_Studies1}

In this section, we illustrate the performance of our proposed sensitivity analysis framework via simulation studies. This section has two goals; one is to illustrate that the blended analysis provides valid estimation when the sensitivity function and parameters are correctly specified, and the other is to illustrate how the blended analysis results are affected by misspecified sensitivity parameters. To achieve these goals, we consider several data-generating scenarios that mimic the motivating data example with MNAR missing values. A data set consists of $N = 1,000$ patients whose records are measured across a one-year window. If there were no missing values, each patient has four measurements: a baseline binary treatment $X$, time-dependent covariates $[Z_1, Z_2]$ where $Z_1$ measured at baseline and $Z_2$ measured at 8 months after baseline but before the measurement of outcome, and a continuous outcome variable $Y$ measured after 1 year from the baseline. The value of $Z_2$ is affected by $X$, and the outcome $Y$ is affected by $X$ and $Z_2$. We are interested in estimating the linear association between baseline covariates $[X, Z_1, XZ_1]$ and the outcome variable $Y$, so we employ ordinary linear regression as the analysis model.

\begin{align}\label{simul_analysismodel1}
Y_i=\beta_0+\beta_1X_i+\beta_2Z_{i1}+\beta_3X_iZ_{i1}+\epsilon_i,\quad\epsilon_i\sim N(0,\sigma^2),\quad i=1,\ldots,1000.
\end{align}

In the simulation study, the baseline covariates $[X, Z_1]$ are fully observed for all patients, but $Z_2$ and $Y$ are subject to missingness due to two reasons: patients who disenroll from the study after 8 months of its initiation will have missing values in their $Z_2$ and $Y$. In addition, patients still enrolled in the study may have missing values in $Z_2$ or $Y$, or both. Consequently, the simulation data provenance can be modularized via three sub-mechanisms $[R_1, R_2, R_3]$. $R_{i1}$ represents whether a patient $i$ is enrolled in the study 8 months after initiation. $R_{i2}$ denotes whether a patient $i$ has time-dependent covariate $Z_2$ measurement, and $R_{i3}$ denotes whether a patient $i$ has the outcome $Y$ measurement. Note that $Z_2$ and $Y$ can only be observed when a patient is enrolled, so $R_1=0$ implies $R_2 = 0$ and $R_3 = 0$. Figure \ref{diag_simul} illustrates the simulation data provenance and three sub-mechanisms.

\begin{figure}[h]
\centering
\includegraphics[height = 7cm]{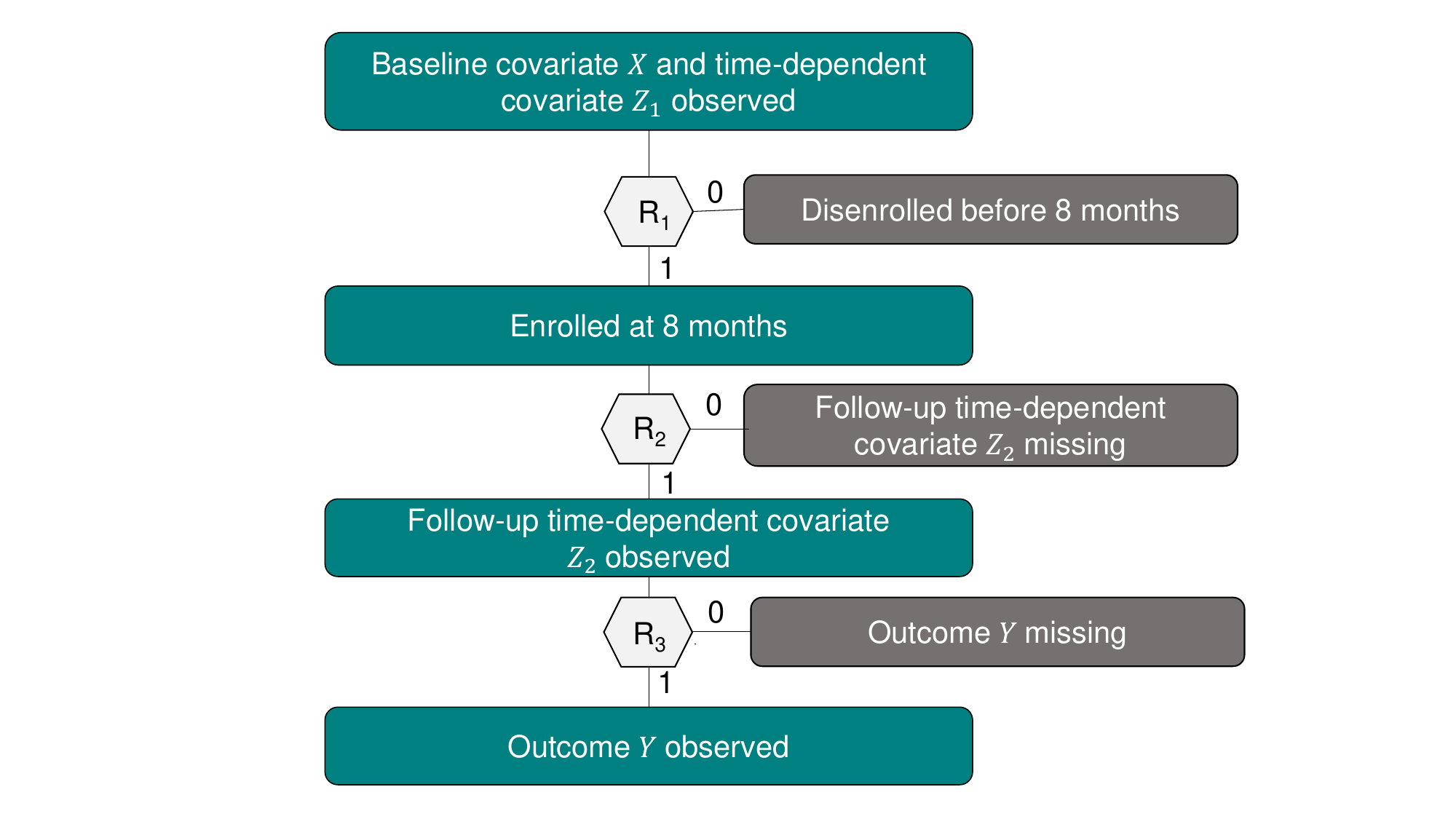}
\caption{A diagram of simulation data provenances}\label{diag_simul1}
\end{figure}

Based on the data provenance, we generate a synthetic data set as follows:
\begin{align}\label{simul_completedata1}
Z_{i1} &\overset{iid}{\sim} Bin(1, expit(-0.5)), \nonumber\\
X_{i1} &\overset{iid}{\sim} Bin(1, expit(-0.5 - 0.25Z_{i1})), \nonumber\\
Z_{i2} &\overset{iid}{\sim} Bin(1, expit(-1.15 - 0.35X_1 + 0.6Z_{i1} + 0.4X_1Z_{i1})), \nonumber\\
Y_{i} &\overset{iid}{\sim} N(0.45 - 0.45X_i + 1.40Z_{i2} - 1.8X_iZ_{i2}, 0.3^2). 
\end{align}

The distribution of $Y$ is directly related to $[X, Z_2]$ only, and thus, its associations with $[X, Z_1, XZ_1]$ are implicit. Consequently, we empirically approximate the true values of analysis model parameters $\boldsymbol\beta$ in Eq. (\ref{simul_analysismodel1}). The approximated analysis model parameters are $\boldsymbol\beta = [0.787, -0.859, 0.176, -0.253]$. In addition, we generate missingness on $[Z_2, Y]$ of each simulated data as 

\begin{align}\label{simul_missingdata1}
\lambda(X, Z_1) &= exp(1.25 + 0.5X - 0.55Z_1 - 0.2XZ_1), \nonumber\\
T&\sim Weibull(\lambda(X, Z_1), 1), \nonumber\\
P(R_1 &= 1\mid X, Z_1) = P(T > 0.68), \nonumber\\
P(R_2 &= 1\mid X, Z_1, Z_2) = expit(1.20 - 0.70X + 0.65Z_1 - 0.55XZ_1 + \delta_{2}Z_2), \nonumber\\
P(R_3 &= 1\mid X, Z_2, Y) = expit(0.45 + 0.35X - 0.70Z_2 - 0.65XZ_2 + \delta_{3}Y),
\end{align}
where $Weibull(\lambda(X, Z_1), 1)$ denotes the Weibull distribution with shape parameter $\lambda(X, Z_1)$ and the scale parameter 1. In such a way, the disenrollment time $T$ is related with baseline covariates $[X, Z_1]$. The hypothetical follow-up time is $0.67$ years ($8$ months), so patients with $T\leq 0.67$ are disenrolled before $Z_2$ is measured, having their $[Z_2, Y]$ as unobserved. In addition, the probabilities of $Z_2$ and $Y$ being unobserved are functions of $[X, Z_1, Z_2]$ and $[X, Z_2, Y]$, respectively. Note that the missing data mechanism is MAR if $\delta_2 = \delta_3 = 0$ and MNAR otherwise. In average, proportion of  $P(R_1 = 1)\approx 0.80$, $P(R_1 = R_2 = 1)\approx 0.57$, and $P(R_1 = R_2 = R_3 = 1)\approx 0.31$ when $\delta_2 = \delta_3 = 0$.

The simulated data provenance consists of three sub-mechanisms and $\tilde{\mathcal{K}} = \{2, 3\}$ because $R_1$ is not related to observing a random variable, and thus IPW (denoted as ``I") is the only possible method, while $[R_2, R_3]$ are open to both IPW or MI (denoted as ``M"). Consequently, there are four possible options of blended analysis: ``III", ``IMI", ``IIM", and ``IMM", depending on how $[\mathcal{K}_{IPW}, \mathcal{K}_{MI}]$ is assigned. For each assignment, we elucidate the blended analysis as follows.

\begin{enumerate}
\item Prespecify sensitivity parameters $[\delta_{z_2}, \delta_{y}]$ for $[R_2, R_3]$.
\item Using all patients, fit a Cox proportional hazard model as $\lambda(t) = \lambda_0(t)exp(\alpha_{1,0}X + \alpha_{1,1}Z_1 + \alpha_{1,2}XZ_1)$ and estimate the baseline survival function $S_0(t)$ using the Breslow estimator \citep{Breslow1972} and the estimated coefficients $[{\hat\alpha}_{1,0}, {\hat\alpha}_{1,1}, {\hat\alpha}_{1,2}]$. The IPW for patients with $R_{i1} = 1$ can be calculated as $w_{1}^{-1} = {\hat P}(T > 0.67 \mid X, Z_{1}) = S_0(0.67)^{exp(\hat\alpha_{1,0}X + \hat\alpha_{1,1}Z_{1} + \hat\alpha_{1,2}XZ_1)}$.
\item If $2\in \mathcal{K}_{IPW}$, we fit a selection model using observations with $R_{1} = 1$ and calculate $w_2$ as follows: 
\begin{align}
w_2^{-1} &= P(R_2 = 1\mid R_1 = 1) = expit(\hat{\alpha}_{2,0} + \hat{\alpha}_{2,1}X + \hat{\alpha}_{2,2}Z_1 + \hat{\alpha}_{2,3}XZ_1 + \delta_{z_2} Z_2).
\end{align}
If $2\in \mathcal{K}_{MI}$, we estimate an imputation model for $Z_2$ using observations with $\bar{R}_{2} = 1$ as follows:
\begin{align}
[Z_2^o \mid \bar{R}_{2} = 1, R_3 = 0] &\sim Bin(1, expit(\hat{\alpha}_{2,0} + \hat{\alpha}_{2,1}X + \hat{\alpha}_{2,2}Z_1 + \hat{\alpha}_{2,3}XZ_1)),\nonumber\\
[Z_2^o \mid \bar{R}_{3} = 1] &\sim Bin(1, expit(\hat{\alpha}_{2,4} + \hat{\alpha}_{2,5}X + \hat{\alpha}_{2,6}Z_1 + \hat{\alpha}_{2,7}XZ_1 + \hat{\alpha}_{2,8}Y + \hat{\alpha}_{2,9}XY)),
\end{align}
Note that the outcome variable $Y$ should be included in the imputation model for $Z_2$ to ensure the unbiased estimation \citep{moons2006using,sterne2009multiple}. Using the estimated imputation model, impute missing values and obtain $Z_2^{(l)} =[Z_2^{o}, Z_2^{m(l)}], l=1,\ldots,M$.
\item If $3\in \mathcal{K}_{IPW}$, we fit a selection model for $Z_2$ using observations with $\bar{R}_{2} = 1$ (i.e., both complete and imputed observations up to $R_2$) as follows:
\begin{align}
w_3^{-1} &= P(R_3 = 1\mid \bar{R}_2 = 1) = expit(\hat{\alpha}_{3,0} + \hat{\alpha}_{3,1}X + \hat{\alpha}_{3,2}Z_2^{(l)} + \hat{\alpha}_{3,3}XZ_2^{(l)} + \delta_{y}Y ).
\end{align}
If $3\in \mathcal{K}_{MI}$, we fit an imputation model for $Y$ using observations with $\bar{R}_{2} = 1$, and draw a random sample $Y^{m}$ for all subjects with $R_{3} = 0$ as follows: 
\begin{align}
[Y^{0}\mid \bar{R}_3 = 1] &\sim N(\hat\alpha_{3,0} + \hat\alpha_{3,1}X + \hat\alpha_{3,2}Z_2^{(l)} + \hat\alpha_{3,2}XZ_2^{(l)},~\hat\alpha_{3,4}^2).
\end{align}
\item Fit an analysis model Eq. (\ref{simul_analysismodel1}) on the observed/imputed data $[X, Z_{1}, Y^{(l)}]$ using $w^{(l)} = \prod\limits_{k=1}^{3}w_{k}^{(l)}$ as a weight for all subjects with $\bar{R}_{3} = 1$ and obtain $\boldsymbol{\hat\beta}_{[\delta_{z_2},\delta_y]}^{(l)}$. The subscript $l$ can be ignored if ``III" is chosen.\item Obtain the final estimates $\boldsymbol{\hat\beta}_{[\delta_{z_2},\delta_y]} = \frac{1}{M}\sum\limits_{l=1}^M\boldsymbol{\hat\beta}_{[\delta_{z_2},\delta_y]}^{(l)}$. The step can be ignored if ``III" is chosen.
\end{enumerate}

In the rest of the section, we present two simulation studies under two different MNAR scenarios. In the Section \ref{mnar_r2}, we discuss the first scenario where we set $[\delta_2, \delta_3] = [0.5, 0.0]$ in Eq. (\ref{simul_missingdata1}) so that missing values on $Z_2$ are MNAR. In the Section \ref{mnar_r32} we illustrate the second scenario where $[\delta_2, \delta_3]$ are $[0.0, 0.5]$; missing data mechanisms on $Y$ are MNAR. For each simulation scenario, we simulate $1,000$ data sets. For each simulated data, we repeat steps $2.\sim 6.$ using various sensitivity parameters and evaluate relative biases (that is, $(\hat\beta - \beta)/|\beta|\times 100\%$) from $1,000$ repetitions at each $[\delta_{z_2}, \delta_{y}]$. In addition, if the relative biases are close to $0\%$, we evaluate coverage probabilities of $95\%$ confidence intervals and see if they achieve nominal coverage.

\subsection{Scenario 1: MNAR on covariate}\label{mnar_r2}

\begin{figure}
\centering
\includegraphics[width = 16cm, height = 11cm]{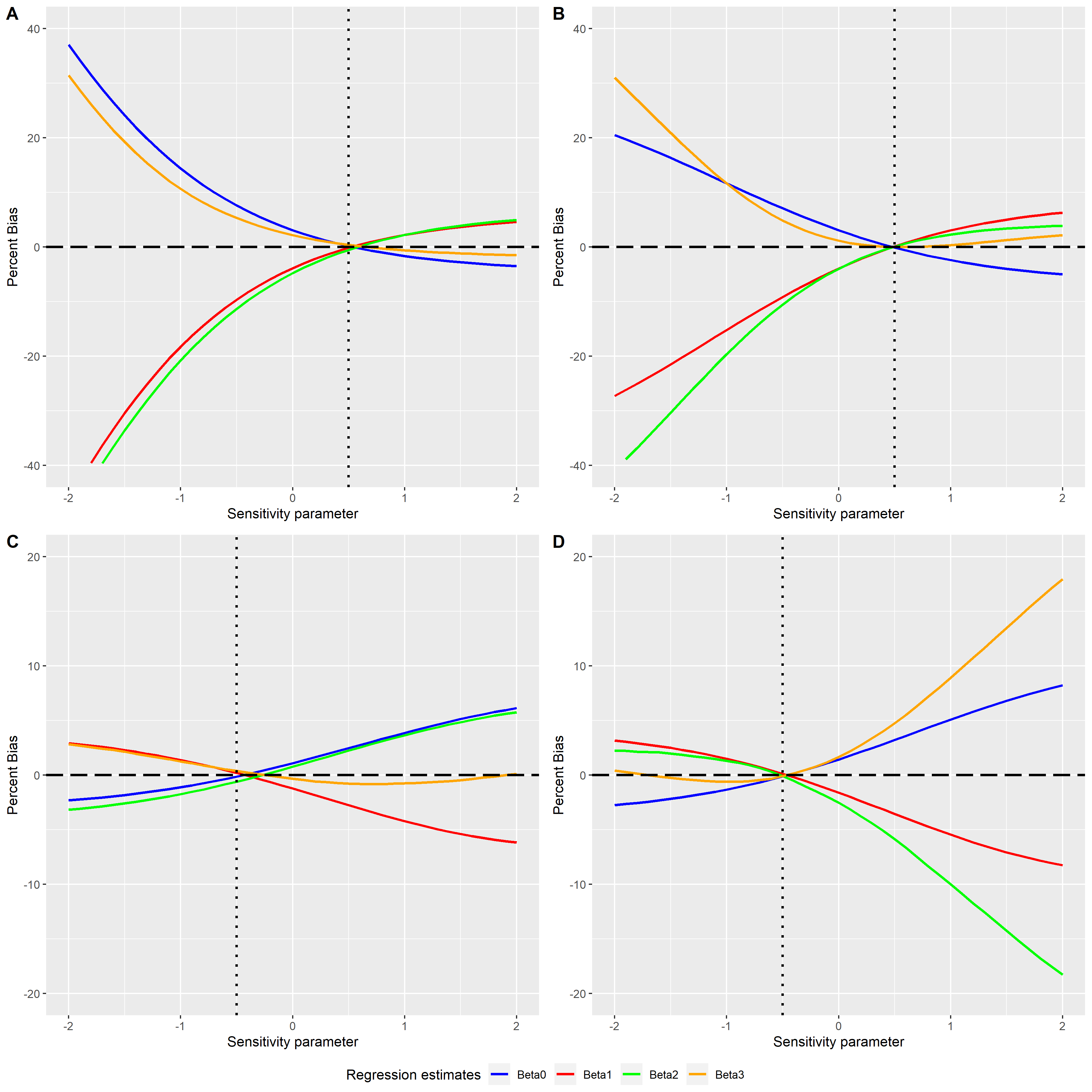}
\caption{Simulation results of Scenario 1. In each plot, the y-axis indicates the percent biases of four regression coefficient estimates, and the x-axis indicates the sensitivity parameter value $\delta_{z_2}$, ranging from $-2.0$ to $2.0$}\label{simulplot_R22}
\end{figure}

In this simulation, we fix $\delta_{y} = 0$ and vary $\delta_{z_2}$ from $-2.0\leq \delta_{z_2}\leq 2.0$ in $0.1$ increments. At each values of $[\delta_{z_2}, \delta_{y}]$, we estimate regression coefficients and evaluate their percent biases. Simulation results are shown in Figure \ref{simulplot_R22}, with one panel for each blended analysis approach. Each individual panel demonstrates how the relative bias changes as a function of $\delta_2$ across the range considered for each of the 4 analysis model parameters $\boldsymbol\beta_{\delta_{z_2}} = [\beta_0, \beta_1, \beta_2, \beta_3]_{\delta_{z_2}}$.  

The top two panels concern ``III" and ``IIM", where missingness in $Z_2$ are addressed via inverse probability weighting. Bias in estimation of all regression coefficient estimates are close to $0\%$ at $\delta_{z_2} = 0.5$, which is  expected because both the sensitivity function and parameter are correctly specified. Similarly, the bottom two panels show relative bias in as a function of $\delta_2$ for ``IMI" and ``IMM", where missingness in $Z_2$ is addressed via multiple imputations. Bias is eliminated at $\delta_{z_2} = -0.4$ for ``IMI" and at $\delta_{z_2} = -0.6$ for ``IMM". We denote $\delta_{Z_2}^*$ as the debiasing sensitivity parameter value for each blended analysis. For all 4 blended analysis procedures, the Bootstrap-MI approach resulted in reasonable coverage for all covariates at $\delta_{z_2}^*$. 

As $\delta_{z_2}$ deviates from $\delta_{Z_2}^*$, the relative bias for all regression coefficients diverges from $0$. Generally, bias was greater when $\delta_{z_2} < \delta^*$ than $\delta_{z_2} \geq \delta^*$ when ``III" and ``IIM" are used, while the opposite trend was observed under ``IMI" and ``IMM". Dramatic bias observed when $\delta_{z_2} < \delta_{z_2}^*$ for  ``III" and ``IIM" is due to the fact that selection probability $P(R_2 = 1\mid X, Z_1^{(l)})$ decreases toward $0$, leading to very large weights $w_2^{(l)}$. 

When a sub-mechanism is addressed via MI, it is not obvious \textit{a priori} which value of $\delta_{z_2}$ will eliminate the bias. This is because the true missingness mechanism is based on a selection model, while imputing under MNAR utilizes a pattern-mixture model-based approach which specifies how the distribution of $Z_2$ differs when it is observed compared to when it is missing. Still, we observe that bias is resolved when $\delta_{z_2} = -0.4$ under ``IMI" and $\delta_{z_2} = -0.6$ under ``IMM". This corresponds to the nature of the missingness mechanism, as $Z_2$ being more likely to be missing when it is 0 than 1 implies that missing values of $Z_2$ are more likely to be 0 than 1, and so the probability of imputing $Z_2$ as $1$ should be adjusted downward.

\subsection{Scenario 2: MNAR on outcome}\label{mnar_r32}

\begin{figure}
\centering
\includegraphics[width = 16cm, height = 11cm]{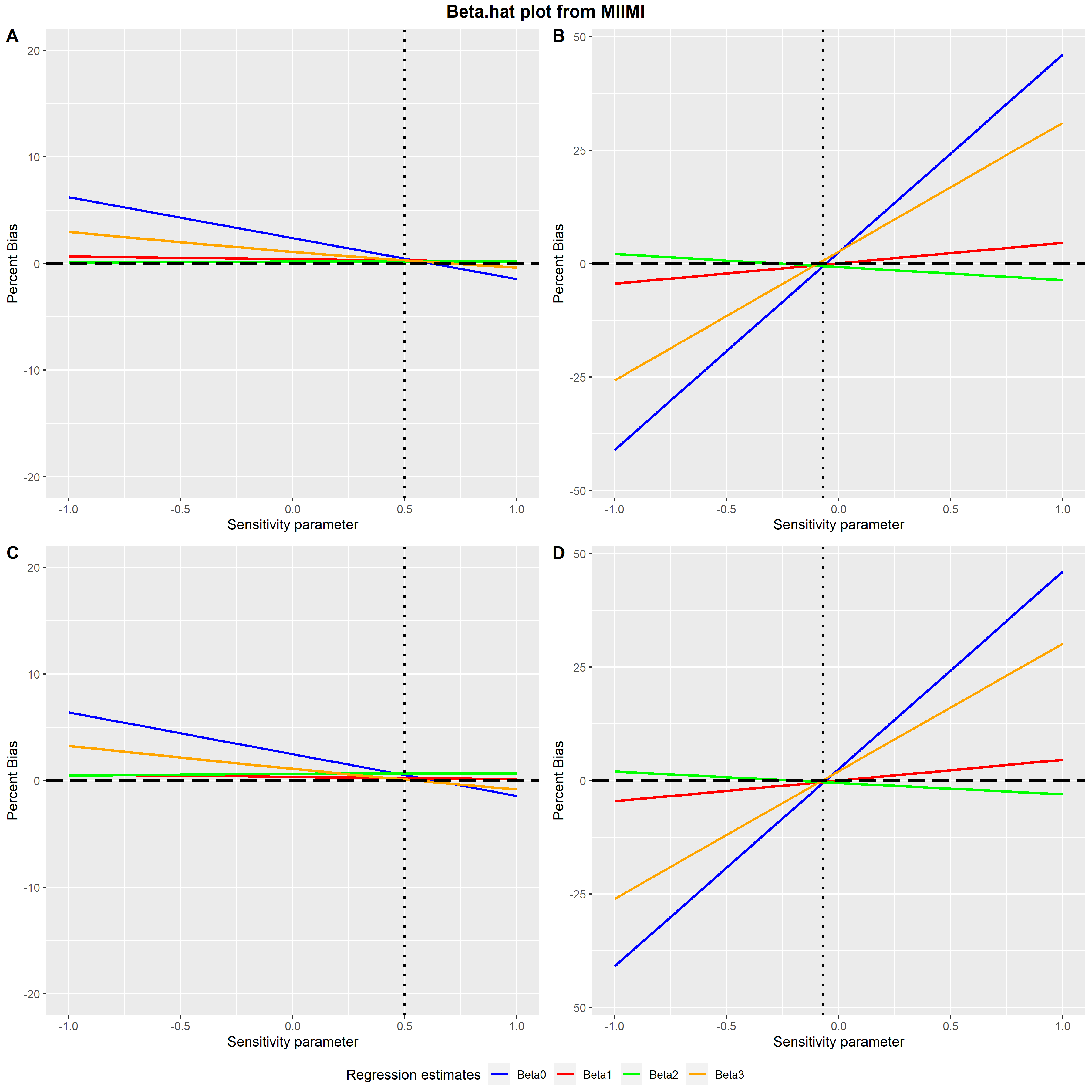}
\caption{Simulation results of Scenario 2. In each plot, the y-axis indicates the percent biases of four regression coefficient estimates, and the x-axis indicates the sensitivity parameter value $\delta_y$, ranging from $-1.0$ to $1.0$}\label{simulplot_R32}
\end{figure}

In this scenario, we fix $\delta_{z_2} = 0$ and vary $\delta_{y}$ from $-1.0\leq\delta_{y}\leq 1.0$ in increments of $0.1$. At each values of $[\delta_{z_2}, \delta_{y}]$, we estimate regression coefficients and evaluate their percent biases. Results are shown in Figure \ref{simulplot_R32}. In ``III" and ``IMI", missingness in $Y$ is addressed via inverse probability weighting (left two panels). We observe that the percent biases of regression estimates are almost completely eliminated when $\delta_{y} = 0.5$. Such a phenomenon is as anticipated, because $\delta_{y} = 0.5$ reflects the true missing data mechanism of outcome variable.

Similarly, the right two panels correspond to ``IIM" and ``IMM", where missingness in $Y$ is addressed via MI. Biases was almost completely removed at $\delta_{y} = -0.08$ for both approaches. Similar to Scenario 1, the particular value of $\delta_y$ that resolve the bias is not immediately obvious from the data generating mechanism and may differ if an alternative sensitivity function is used. Still, it correctly reflects that a smaller values of $Y$ are more likely to be missing, and thus, it is reasonable that imputing smaller values of $Y$ than you would based on the observed data would resolve the bias. Finally, the coverage probabilities of the $95\%$ Bootstrap-MI confidence intervals are also close to $0.95$ in all cases with appropriate $\delta_{y}$. 

\section{Strategies for choosing ranges of sensitivity parameters}

One strategy of choosing the ranges of each sensitivity parameter is to based on its implication on the corresponding selection probability or imputation model. When a sub-mechanism $R_k$ is addressed via MI, the average imputed values of $D_k^m$ can be used to explicitly understand the effect of a particular $\delta_k$ on the analysis. For example, if the blended analysis is implemented under ``IMIIM", $\delta_2$ and $\delta_5$ represent the additive increase in average imputed BMI at baseline and follow-up, respectively. The average imputed baseline BMI at $\delta_2 = 0$ is $46.178$, and moves between $[40.178, 52.178]$ as $\delta_2$ moves between $[-6.0, 6.0]$. Based on such implications of a particular $\delta_2$ on unobserved baseline BMI, the analyst may specify appropriate ranges of sensitivity parameters to properly reflect the MNAR missing data mechanism of concern.

When IPW is used for $R_k$, implications of a particular $\delta_k$ on the blended analysis result can be explicitly evaluated via the distribution of $D_k^m$. For example, when $R_1$ is modeled via IPW, the conditional probability $P(CKD_0 = 1 \mid R_1 = 0, \delta_1)$ reflects the implication of $\delta_1$ on the distribution of unobserved baseline CKD. Although this quantity is generally unobserved, it can be identified from the observed data and the selection probability specified from the blended analysis. Specifically, we have assumed that $P(R_1 = 1\mid CKD_0 = 1, {\bf X}, \delta_1) = expit({\bf X}\boldsymbol\alpha + \delta_1CKD_0)$ when ``IMIIM" is used, which can be used for identifying a subject-level conditional probability $P(CKD_0 = 1 \mid R_1 = 0, {\bf X}, \delta_1)$ as follows:

\begin{align}\label{eq1}
P(CKD_0 = 1 \mid R_1 = 0, {\bf X}, \delta_1) &= \frac{P(R_1 = 0, CKD_0 = 1\mid {\bf X},\delta_1)}{P(R_1 = 0\mid {\bf X},\delta_1)} \nonumber\\
&= \frac{P(R_1 = 0\mid CKD_0 = 1, {\bf X}, \delta_1)P(CKD_0 = 1\mid {\bf X},\delta_1)}{P(R_1 = 0\mid {\bf X},\delta_1)} \nonumber\\
&= \frac{P(R_1 = 0\mid CKD_0 = 1, {\bf X}, \delta_1)\frac{P(R_1 = 1, CKD_0 = 1\mid {\bf X}, \delta_1)}{P(R_1 = 1\mid CKD_0 = 1, {\bf X},\delta_1)}}{P(R_1 = 0\mid {\bf X}, \delta_1)} \nonumber\\
&= \frac{P(R_1 = 0\mid CKD_0 = 1, {\bf X}, \delta_1)P(R_1 = 1, CKD_0 = 1\mid {\bf X}, \delta_1)}{P(R_1 = 1\mid CKD_0 = 1, {\bf X}, \delta_1)P(R_1 = 0\mid {\bf X}, \delta_1)}.
\end{align}

Based on the subject-level probability, we can describe the conditional distribution of unobserved $CKD_0$ as follows:
\begin{align}\label{eqx}
P(CKD_{0} = 1 \mid R_{1} = 0, \delta_1) &= \int P(CKD_{0} = 1 \mid {\bf X}, R_{1} = 0, \delta_1)dF({\bf X}\mid R_1 = 0, \delta_1) \nonumber\\
&=\int \frac{P(R_{1} = 0\mid CKD_{0} = 1, {\bf X}, \delta_1)}{P(R_{1} = 1\mid CKD_{0} = 1, {\bf X}, \delta_1)}\frac{P(R_{1} = 1, CKD_{0} = 1\mid {\bf X}, \delta_1)}{P(R_{1} = 0\mid {\bf X}, \delta_1)} dF({\bf X}\mid R_1 = 0, \delta_1). 
\end{align}

An empirical estimator of Eq. (\ref{eqx}) is
\begin{align}\label{eqx2}
{\widehat P}(CKD_{0} = 1 \mid R_{1} = 0, \delta_1)  &= \frac{1}{n_{\mathcal{R}_2}}\sum\limits_{i=1}^Nexp(-{\bf X}_i^{'}\boldsymbol\hat\alpha_1 - \delta_1)\frac{{\widehat P}(R_{i1} = 1, CKD_{i0} = 1\mid {\bf X}_i)}{{\widehat P}(R_{i1} = 0\mid {\bf X}_i)}I(R_{i1} = 0),
\end{align}
where $\mathcal{R}_{CKD_0}$ denotes patients with $CKD_{0} = 1$, $n_{\mathcal{R}_2}$ denotes the number of patients with $R_{i1} = 0$, and $\boldsymbol\hat\alpha_1$ is the regression coefficient estimates related to ${\bf X}$ in the IPW function. A patient-level probability estimates ${\widehat P}(R_{i1} = 1, CKD_{i0} = 1\mid {\bf X}_i)$ in Eq. (\ref{eqx2}) can be estimated via a logistic linear model regressing $R_{1}\times CKD_{0}$ on ${\bf X}$. Similarly, ${\widehat P}(R_{i1} = 1\mid {\bf X}_i)$ can be estimated separately, using a logistic linear model regressing $R_{1}$ on ${\bf X}$. Also, Eq. (\ref{eqx2}) assumes that the distribution of $[{\bf X}\mid R_1 = 0]$ is invariant to $\delta_1$.

For another example, when the ``MIIMI" option is used for the blended analysis, $R_2$ is modeled via IPW. In this case, a quantity of interest is $E(BMI_0\mid R_2 = 0, \delta_2)$, the average of the unobserved baseline BMI. To calculate the average, the conditional distribution of $[BMI_0\mid R_2 = 0]$ should be specified. This quantity is generally not identifiable, but we can identify it using the selection probability used for $R_2$ from the blended analysis as 
\begin{align}\label{bmi1}
f(BMI_0 \mid R_{i2} = 0, {\bf X}, \delta_2) &=\frac{P(R_{i2} = 0\mid BMI_0, {\bf X}, \delta_2)}{P(R_{i2} = 1\mid BMI_0, {\bf X}, \delta_2)}\frac{P(R_{i2} = 1\mid {\bf X}, \delta_2)}{P(R_{i2} = 0\mid {\bf X}, \delta_2)}f(BMI_0\mid R_{i2} = 1, {\bf X}, \delta_2),
\end{align}
where $f(BMI_0 \mid R_{i2} = 1, {\bf X}, \delta_2)$ denotes the condtional density function of observed baseline BMI given ${\bf X}$. Note that $f(BMI_0 \mid R_{i2} = 1, {\bf X}, \delta_2)$, the conditional density function of unobserved baseline BMI given ${\bf X}$, is identifiable due to the selection probability $P(R_{i2} = 1\mid BMI, {\bf X}, \delta_2)$ and thus subject to the choice of $\delta_2$. Consequently, the marginal conditional mean of the unobserved $BMI_0$ can be calculated as

{\small
\begin{align}\label{bmi2}
E(BMI_0 \mid R_{i2} = 0, {\bf X}, \delta_2) &= \int BMI_0dF(BMI_0\mid R_{i2} = 0, {\bf X}, \delta_2) \nonumber\\
 &= \int BMI_0\frac{P(R_{i2} = 0\mid BMI_0, {\bf X}, \delta_2)}{P(R_{i2} = 1\mid BMI_0, {\bf X}, \delta_2)}\frac{P(R_{i2} = 1\mid {\bf X}, \delta_2)}{P(R_{i2} = 0\mid {\bf X}, \delta_2)}P(BMI_0\mid R_{i2} = 1, {\bf X}, \delta_2)d(BMI_0), \nonumber\\
E(BMI_0 \mid R_{i2} = 0, \delta_2) &=E_{{\bf X}\mid R_{2} = 0}\left[E(BMI_0 \mid R_{i2} = 0, {\bf X}, \delta_2)\right].
\end{align}
}

Consequently, an empirical estimator of Eq. (\ref{bmi2}) can be obtained as
{\small
\begin{align}\label{bmi3}
{\widehat E}(BMI_0 \mid R_{2} = 0, \delta_2) &=\frac{1}{n_{\mathcal{R}_2}}\sum\limits_{i=1}^N{\widehat E}\left(BMI_{i0}\frac{{\widehat P}(R_{i2} = 0\mid BMI_{i0}, {\bf X}_i, \delta_2)}{{\widehat P}(R_{i2} = 1\mid BMI_{i0}, {\bf X}_i, \delta_2)}\frac{{\widehat P}(R_{i2} = 1\mid {\bf X}_i)}{{\widehat P}(R_{i2} = 0\mid {\bf X}_i)}\mid {\bf X}_i, R_{i2} = 0\right)I(R_{i2} = 0) \nonumber\\
&=\frac{1}{n_{\mathcal{R}_2}}\sum\limits_{i=1}^N{\bf X}_i\boldsymbol{\hat\alpha}_2I(R_{i2} = 0), 
\end{align}
}
where $n_{\mathcal{R}_2}$ denotes the number of patients with $R_{i2} = 0$, and $\boldsymbol{\hat\alpha}_2$ is the regression estimates obtained using ordinary linear model regressing $\frac{BMI_0{\widehat P}(R_{2} = 0\mid BMI_{0}, {\bf X}, \delta_2)}{{\widehat P}(R_{2} = 1\mid BMI_{0}, {\bf X}, \delta_2)}\frac{{\widehat P}(R_{2} = 1\mid {\bf X})}{{\widehat P}(R_{2} = 0\mid {\bf X})}$ on ${\bf X}$. Similar to Eq. (\ref{eqx2}), a patient-level probability estimates ${\widehat P}(R_{i2} = 1\mid BMI_{i0}, {\bf X}_i, \delta_2)$ is directly obtained from the blended analysis, while ${\widehat P}(R_{i2} = 1\mid {\bf X}_i)$ should be calculated separately via logistic regressions. Also, Eq. (\ref{bmi3}) assumes that the distribution of $[{\bf X}\mid R_2 = 0]$ is invariant to $\delta_2$. 

\end{document}